\documentclass[11pt,letterpaper]{article}
\pdfoutput=1
\usepackage{jheppub}

\usepackage{amssymb,amsmath,bm}
\usepackage{color}
\usepackage{slashed}
\usepackage[latin1]{inputenc}
\usepackage{amsfonts}
\usepackage{lscape}
\usepackage{amsthm}
\usepackage{booktabs}
\usepackage{array}
\usepackage{rotating}
\usepackage{float}
\usepackage{multirow}
\usepackage{graphicx,rotating,amsmath,multirow,xcolor,colortbl,xcolor}
\usepackage{hyperref,pdflscape,slashed}
\definecolor{rosso}{cmyk}{0,1,1,0.4}
\definecolor{rossos}{cmyk}{0,1,1,0.55}
\definecolor{rossoc}{cmyk}{0,1,1,0.2}
\definecolor{blu}{cmyk}{1,1,0,0.3}
\definecolor{blus}{cmyk}{1,1,0,0.6}
\definecolor{bluc}{cmyk}{1,1,0,0.1}
\definecolor{verde}{cmyk}{0.92,0,0.59,0.25}
\definecolor{verdec}{cmyk}{0.92,0,0.59,0.15}
\definecolor{verdes}{cmyk}{0.92,0,0.59,0.4}
\definecolor{Gray}{gray}{0.95}

\font\tenrsfs=rsfs10 at 12pt
\font\sevenrsfs=rsfs7
\font\fiversfs=rsfs5
\newfam\rsfsfam
\textfont\rsfsfam=\tenrsfs
\scriptfont\rsfsfam=\sevenrsfs
\scriptscriptfont\rsfsfam=\fiversfs
\def\mathscr#1{{\fam\rsfsfam\relax#1}}

\usepackage[small,bf]{caption}
\newcommand{\gev}{{\rm GeV}}
\newcommand{\mev}{{\rm MeV}}
\newcommand{\kev}{{\rm keV}}
\def\Lag{\mathscr{L}}

\newcommand{\lsim}{\stackrel{<}{_\sim}}

\newcommand{\ie}{{\em i.e.}}
\newcommand{\hc}{{\rm h.c.}}

\newcommand{\be}{\begin{equation}}
\newcommand{\ee}{\end{equation}}
\newcommand{\bea}{\begin{eqnarray}}
\newcommand{\eea}{\end{eqnarray}}
\newcommand{\beq}{\begin{equation}}
\newcommand{\eeq}{\end{equation}}
\newcommand{\beqa}{\begin{eqnarray}}
\newcommand{\eeqa}{\end{eqnarray}}

\def\eq#1{eq.~(\ref{#1})}
\def\ord{\mathcal{O}}

\def\trh{T_{\rm RH}}
\def\fig#1{fig.~\ref{#1}}

\begin{document}

\title{Natural Heavy Supersymmetry}
 
\author[a,b]{Brian Batell,}
\author[b]{Gian F. Giudice,}
\author[b]{and Matthew McCullough}

\affiliation[a]{Pittsburgh Particle Physics, Astrophysics, and Cosmology Center,\\
Department of Physics and Astronomy, University of Pittsburgh, USA}
\affiliation[b]{CERN, Theory Division, Geneva, Switzerland}

\emailAdd{brian.batell@cern.ch}
\emailAdd{gian.giudice@cern.ch}
\emailAdd{matthew.mccullough@cern.ch}

\date{\today}

\abstract{We study how, as a result of the scanning of supersymmetry breaking during the cosmological evolution, a relaxation mechanism can naturally determine a hierarchy between the weak scale and the masses of supersymmetric particles. Supersymmetry breaking is determined by QCD instanton effects, in an extremely minimal setup in which a single field drives the relaxation and breaks supersymmetry. Since gauginos are lighter than the other supersymmetric particles by a one-loop factor, the theory is a realisation of Split Supersymmetry free from the naturalness problem.}


\arxivnumber{1509.00834}

\preprint{CERN-PH-TH-2015-215, PITT-PACC-1512}

\maketitle

\section{Introduction}
\label{secintr}

The cosmological relaxation of the electroweak scale~\cite{GKR} (see ref.~\cite{gia} for earlier attempts and refs.~\cite{altri,zohar} for related work) offers an interesting mechanism to deal with the problem of Higgs naturalness. Instead of introducing new dynamics at the weak scale, as conventionally done in other solutions, it gives an explicit realisation of self-organised criticality~\cite{soc}, in which the system is dynamically attracted towards the near-critical condition for electroweak breaking. This situation is achieved with an axion-like~\cite{axion} particle (called relaxion) which, during the cosmological evolution at the inflationary epoch, scans the order parameter of the electroweak phase transition. Once electroweak symmetry is broken, non-perturbative QCD effects give a back-reaction that prevents the relaxion from rolling much further.

By construction, the setup of ref.~\cite{GKR} requires an energy cutoff, which is found to be considerably smaller than many of the new-physics mass scales that are believed to exist in theories more fundamental than the Standard Model (SM). However, the naturalness problem is rather special because it involves physics from all distance scales, no matter how small. If naturalness is solved in an effective theory with cutoff $\Lambda$, hypothetical new particles that live beyond the regime of the effective theory can easily reintroduce an even bigger problem if they couple (directly or indirectly) to the Higgs.  In other words, solving the Higgs naturalness up to a cutoff scale $\Lambda$ is a very important result from the phenomenological point of view but, in a broader perspective, is just a way to postpone the real problem to higher scales. Moreover, the relaxion mechanism is a solution tailored to cure the quantum properties of the Higgs. However, in the broader perspective we want to adopt, one can expect that more fundamental theories will require the presence of other scalar particles than the Higgs, such as the inflaton, GUT-like states, or fields related to dark energy. Any of these scalar particles will introduce their own naturalness problem, and each one will require a solution unrelated to the Higgs relaxation. For these reasons, we claim that the Higgs relaxation mechanism is satisfying in the IR, but cries out for a UV picture.

Supersymmetry offers an elegant solution to the Higgs naturalness problem. From the UV point of view, the solution given by supersymmetry is very ambitious. The naturalness problem is solved not only for the Higgs, but for any scalar particle in the theory and with no cutoff limitation. This makes the supersymmetric framework very attractive for a variety of problems in high-energy physics and cosmology, well beyond the issue of electroweak breaking. Moreover, supersymmetry appears as a necessary field-theoretical link with string theory and, therefore, possibly with quantum gravity. Unfortunately, this magnificent UV picture is not corroborated by IR information. Experiments have not detected the presence of supersymmetry up to about the TeV scale, while a resolution of the Higgs naturalness would require supersymmetric particles with masses around $M_Z$. In summary, supersymmetry gives a splendid UV picture, but suffers in the IR.

These considerations lead us to believe that relaxation and supersymmetry could be a perfect match. Supersymmetry will deal with the grand picture of particle physics, while the relaxation mechanism will explain the so-called ``little hierarchy problem", {\it i.e.} the separation between the scales of supersymmetry and electroweak breaking. This is the theoretical framework that we want to explore in this paper.

To achieve this goal, we treat the relaxion as a QCD-like axion, except for the unfamiliar property that the field spans a non-compact space. This hypothesis is a strong departure from the common interpretation of the Goldstone modes as excitations around a compact field space and its realisation requires going beyond the ordinary rules of quantum field theory.\footnote{As this paper was being completed, ref.~\cite{zohar} appeared in which it was claimed that, within quantum field theory, the cutoff of any UV completion realising the relaxion as an axion must be around the weak scale and the small parameter in the relaxion potential is not natural. These conclusions put on firmer grounds the widespread belief that the non-compact properties of the relaxion must originate from physics beyond quantum field theory. See section~\ref{secUV} for more comments on this point.}  However, fields with such unusual properties have been conjectured to exist in the context of string theory and the underlying mechanism is monodromy~\cite{monod}. As the field winds around its periodic potential, its energy increases at each cycle due to its couplings to fluxes. This effectively allows for large super-Planckian excursions of the field. In our context, we assume that the shift symmetry of the relaxion is broken by a small parameter that generates a sliding potential. From a field-theory point of view, the smallness of this parameter is natural according to 't Hooft criterion~\cite{hooft}, although a final assessment requires knowledge of its non-field-theoretical origin. 

At the beginning of inflation, the relaxion is found very far from its true vacuum, and thus the field starts slow-rolling towards the minimum of the potential. During this evolution, the vacuum energy associated with the relaxion changes. This vacuum energy breaks supersymmetry and is the leading source of soft terms for the partners of the SM particles. This means that the soft terms effectively scan during the history of the universe. When the soft terms become of the order of the supersymmetric mass parameter $\mu_0$, the symmetric vacuum of the Higgs potential is suddenly destabilised and electroweak symmetry is spontaneously broken. This triggers a back-reaction on the relaxion potential from non-perturbative QCD effects, which stops the field from further evolution. As a result, the Higgs vacuum expectation value is found near the critical condition for electroweak breaking, while the soft masses are $\ord (\mu_0 )$. Note that the value of $\mu_0$ is not correlated with the weak scale in the fundamental theory. Nonetheless, the hierarchy between $\mu_0$ and the weak scale is not the result of a tuning, but of a dynamical relaxation mechanism.

In combining relaxation and supersymmetry, we find that the total is much greater than the sum of the two parts. New interesting elements emerge, which were not evident in the individual theories. From the point of view of the relaxion, we have gained a controllable UV completion, which tames any possible contribution from physics above the cutoff that could spoil Higgs naturalness. Indeed, the parameter $\mu_0$ (which is the typical size of the soft terms) plays the role of the cutoff in the setup of ref.~\cite{GKR}. Moreover, the scanning of the supersymmetry breaking scale is an automatic feature of any theory in which the field varies. This is because the breaking of global supersymmetry is associated with the vacuum energy of the theory. Whenever the relaxion slow-rolls, the scale of supersymmetry breaking (and, consequently, the order parameter for electroweak breaking) scans. The mechanism does not require any special interactions between the relaxion and other fields that violate the shift symmetry and are not accounted for by monodromy. On the contrary, such interactions are needed in the case of ref.~\cite{GKR} where it is necessary to introduce a PQ-violating coupling between the Higgs and the relaxion. 

From the point of view of supersymmetry, we have gained a natural explanation of the little hierarchy problem, which was the original target of this work. But on the way, we have also discovered an economical and elegant way of breaking supersymmetry. The theory is extremely economical in terms of field content. Besides the usual SM superfields, we have added only one chiral superfield. The pseudoscalar component of this supermultiplet is the relaxion, whose scanning value breaks supersymmetry; the fermionic component is the Goldstino. Supersymmetry is broken in a metastable vacuum generated by the interplay between the minute PQ-breaking effect and QCD instantons. The essential simplicity of the supersymmetry-breaking structure makes the framework interesting, quite independently of the relaxation mechanism. The basic reason for this simplicity can be traced back to the general idea of breaking supersymmetry in metastable vacua~\cite{iss}, as a way to avoid the theorems that dictate very constraining conditions~\cite{theorms} on supersymmetry-violating absolute minima. In our theory, supersymmetry is recovered at the bottom of the relaxion potential, but non-perturbative QCD effects trap the field very far from its true vacuum. The main obstacle for the viability of the theory is the strong CP problem. Its resolution requires some modifications of the minimal model and we present some possible ways out.

By construction our relaxation mechanism predicts that the supersymmetric particles are heavy, with masses parametrically unrelated to the weak scale. Nonetheless, constraints from inflationary dynamics imply that their masses must be smaller than some hundreds of TeV. An interesting feature of our setup is that gaugino masses are smaller than squark mass by a one-loop gauge factor. This makes the spectrum very similar to Mini-Split models~\cite{minisplit} emerging from anomaly mediation, and gives some hope of detection at the LHC and, especially, at future colliders operating in the 100 TeV domain. An important difference of our scenario is that, unlike the case of anomaly mediation, the gravitino is fairly light. As a result, we find some very characteristic signatures at hadron colliders. 

\section{The framework}
\label{secframe}

Our theoretical setup is simple and minimal. We consider an effective theory valid below the PQ symmetry breaking scale $f$, in which the only degrees of freedom are the usual fields of the supersymmetric extension of the SM together with a new chiral superfield $S$. The superfield $S$ describes the relaxion ($a$), its scalar counterpart (the srelaxion $s$), and its supersymmetric partner (the relaxino $\tilde a$),
\beq
S  =  \frac{s+i\,a}{\sqrt{2}} + \sqrt{2}\, \theta\, \tilde a+ \theta^2\, F +~{\rm derivative~terms}.
\eeq
For convenience, we choose $S$ to be dimensionless. The transformations under PQ symmetry of $S$ and of the quark, lepton and Higgs chiral superfields (collectively denoted as $\Phi_i$) are
\bea
S & \rightarrow & S +i\, \alpha 
\label{transS} \\
\Phi_i & \rightarrow & e^{i q_i \alpha} \, \Phi_i,
\label{transPh}
\eea
where $q_i$ are the PQ charges and $\alpha$ is the global transformation parameter. From \eq{transS} we see that, under PQ transformations, the relaxion changes by a shift ($a\to a+\sqrt{2}\, \alpha$), while $s$ and $\tilde a$ remain invariant. We assign the PQ charges such that the Yukawa interactions are invariant, but we allow for the possibility that the gauge-invariant Higgs bilinear carries a PQ charge
\beq
H_uH_d \to e^{i q \alpha} \, H_uH_d\, , \quad \quad \quad q\equiv q_{H_u}+q_{H_d} \, .
\eeq
The case $q=0$ belongs to the class of KSVZ~\cite{ksvz} axion models, in which the PQ sector is made of heavy matter, while the case $q\ne 0$ describes DFSZ~\cite{dfsz} models, in which the ordinary Higgs fields are charged under PQ.

The most general Lagrangian, up to dimension-4 interactions invariant under supersymmetry and PQ, is given by\footnote{For the effective theory of the supersymmetric axion, see~\cite{revax} and references therein.}
\bea
\Lag &=& \int d^4 \theta \left[ f^2 K(S+S^\dagger) +Z_i(S+S^\dagger)\, \Phi_i^\dagger e^V \Phi_i \right]
+ \left[ \int d^4 \theta \, U(S+S^\dagger)\,  e^{-qS}H_uH_d  \right. \nonumber \\
&+& \left. \int d^2\theta \Big( C_a(S) \, {\rm Tr} {\cal W}_a{\cal W}_a + \mu_0\, e^{-qS}H_uH_d+{\rm Yukawa~int.} \Big) +\hc \right] \, ,
\label{lagg}
\eea
\beq
C_a(S) = \frac{1}{2g_a^2} -\frac{i\, \Theta_a}{16\pi^2} -\frac{c_a\, S}{16\pi^2} \, .
\label{defanomal}
\eeq 
Here the index $a$ runs over the 3 factors of the SM gauge group and $K$, $Z_i$, $U$ are generic functions of the combination $S+S^\dagger$ (which contains $a$ only through derivative terms).\footnote{The factor $e^{-qS}$ in \eq{lagg} can be eliminated by a superfield redefinition, as discussed in appendix A.}
 
At this stage, the potential for $a$ exactly vanishes because of the shift symmetry, while supersymmetry insures that $s$ and $\tilde a$ remain massless too. To obtain a non-trivial dynamical evolution of the relaxion we introduce an explicit breaking of the shift symmetry that mimics the effect of monodromy. We choose to break softly the shift symmetry through a small mass parameter $m$ (with $m\ll f$) in the superpotential $W$

\beq
W/f^2=\frac{m}{2} \, S^2\, .
\label{polpot}
\eeq
For simplicity, we take $m$ real. At the field-theory level, the hypothesis $m\ll f$ is technically natural because the theory acquires a larger symmetry in the limit $m\to 0$. We have chosen a superpotential quadratic in $S$ but, as we will show in section~\ref{secUV}, any other form of $W$ would lead to the same conclusions. Of course, a superpotential linear in $S$ is not useful because no potential for $a$ is generated.

With the inclusion of the term in \eq{polpot}, the Lagrangian for the relaxion multiplet at zero derivatives is
\beq
\Lag / f^2 = \kappa^{-1}(s) F^*F + m \left[ \left( \frac{s+i\,a}{\sqrt{2}} \right) F + \hc \right] \, 
\eeq
where $\kappa(s)=1/K'' (\sqrt{2} s)$ and $K''$ is the second derivative of the K\"ahler function in \eq{lagg}.
For instance, if $K$ were approximately canonical at small field value, \ie\ $K=(S+S^\dagger)^2/2 +  
\ord [(S+S^\dagger)^3]$, then $\kappa(s)=1+\ord (s)$.
Solving the equation of motion for the auxiliary field $F$, we find
\beq
F =-m  \left( \frac{s-i\,a}{\sqrt{2}} \right) \kappa (s) \, .
\label{auxil}
\eeq
From this we obtain the scalar potential for $a$ and $s$
\beq
V/f^2 =\frac{m^2}2 ( s^2+a^2 )\kappa (s) \, .
\label{che}
\eeq 
 
The potential in \eq{che} has a supersymmetry-preserving minimum at $a=0$, $s=0$. However, we assume that at the beginning of inflation $a$ is displaced far from its minimum and starts at a value $a\gg 1$. 

On a fixed $a$ background, the potential in \eq{che} is minimised at $s=\bar s$, with $\bar s$ given by the solution of the equation $\kappa' (\bar s ) \approx 0$ (valid in the limit $a\gg 1$). Here we are making the assumption that the function $\kappa$ is positive and generic; hence $\bar s = \ord (1)$. We will not need to know the exact location of $\bar s$, but the important point is that $\bar s$ does not depend on $a$ in the limit $a\gg 1$. As a result, $\bar s$ will not change during the cosmological evolution of the relaxion, as long as $a$ scans very large field values. 

Our assumption that in the early universe $a$ starts at a large field value, while $s$ and all scalar fields of the supersymmetric SM lie at the minimum of the potential, can be viewed as self-consistent. As we will show in the following, on the relaxion background, all scalar fields other than $a$ acquire masses larger than the Hubble rate, and therefore it is natural to expect that, at the beginning of inflation, they are found at their minimum. This is not necessarily true for the relaxion field.

On the relaxion background, the relaxion, srelaxion, and relaxino masses and the auxiliary field are proportional to
\beq
m_a \propto m\, , \quad \quad m_s \propto m a\, , \quad \quad 
m_{\tilde a} \propto m\, , \quad \quad
F \propto m a\, .
\eeq

Here we have omitted factors of order unity coming from wave-function renormalisation. The important point is that, during the cosmological evolution in the range $a\gg 1$, the mass of the srelaxion and the supersymmetry-breaking scale scan linearly with $a$ and are much larger than the curvature of the quadratic potential on which $a$ rolls.

Since the relaxion background breaks supersymmetry, effective soft terms are generated, as in axion-mediation \cite{jmr}. The complete calculation of the soft terms is presented in appendix A. Here we give only approximate expressions that exhibit the parametric dependence. Gauginos acquire their masses from the coupling between $S$ and the gauge field strength $\cal W$ in the superpotential of \eq{lagg},
\beq
M_{{\tilde g}_a} =\frac{g_a^2 c_a F}{16\pi^2} \approx \frac{\alpha_a}{4\pi}\, ma \, .
\label{pippog}
\eeq
Gaugino masses are expected to be a one-loop factor smaller than the parametric scale of supersymmetry breaking $F\approx ma$.\footnote{Here we have derived $F\approx ma$ as an approximate relation, but we suspect that this equation must have a more universal validity. However, we expect that effects suppressed by powers of the Planck mass (neglected here) will modify our equation.} 

Soft scalar masses are induced by the functions $Z_i$ in the K\"ahler potential of \eq{lagg}. However, while the coupling between $S$ and gauge superfields is determined by the anomaly condition, the coupling between $S$ and matter is more model-dependent. For this reason, we introduce a new mass scale $M_*$ (with $M_* \ge f$) that parametrizes the mediation of direct couplings between matter and the relaxion superfield through the K\"ahler interaction
\beq
\Lag = \frac{f^2}{M_*^2} \int d^4 \theta \, (S+S^\dagger)^2 \Phi_i^\dagger \Phi_i  \, .
\eeq
This gives scalar soft masses parametrically equal to
\beq
{\tilde m}_i \approx \frac{f}{M_*}\, ma\, .
\eeq
If $M_* \approx f$, as expected in the most general effective theory, then scalar masses dominate the supersymmetric spectrum, with gauginos lighter by a one-loop factor. However, making use of the non-renormalisation theorem of supersymmetry, it is possible to imagine setups where $M_*$ is much larger than $f$. If $M_* \gg 4\pi f/ \alpha$, gaugino masses are the dominant source of supersymmetry breaking in the visible sector. The latter case occurs, for instance, for gravity mediation ($M_*\approx M_P$) and $f \lsim 10^{16}~\gev$.

Soft-breaking trilinear couplings of order $A_{ijk}\approx ma$ are generated for general functions $Z_i$, but could be suppressed by powers of $f/M_*$ if the mediation between the matter and relaxion sector occurs only through heavy states. However, whatever value of $M_*$ is chosen, trilinear couplings, scalar and gaugino masses scale linearly with $a$. 

Since supersymmetry allows for a $\mu$ term in the Lagrangian, the scaling of $\mu$ with $a$ is different than in the case of scalar and gaugino masses and this will be crucial for our relaxation mechanism. 
For phenomenological reasons, we are interested in taking the mass parameter $\mu_0$ in \eq{lagg} much smaller than the PQ scale $f$. The non-renormalisation property of the superpotential insures that the condition $\mu_0 \ll f$ is technically natural. The hierarchy between these two masses is just a reincarnation of the usual $\mu$-problem in our context.

Besides the explicit $\mu_0$ in \eq{lagg}, there are supersymmetry-breaking sources for $\mu$ and $B_\mu$ (which is the coefficient of the scalar $H_uH_d$ bilinear in the potential). The complete calculation of these terms is presented in appendix A. Here we only show the parametric dependence:
\beq
\mu =\mu_0- c_\mu \, ma \, ,~~~~~
B_\mu=c_0\, \mu \, ma+c_B  \, m^2a^2 \, .
\label{mubmu}
\eeq
We find that $\mu$ is given by the sum of two contributions. The first one is $\mu_0$, which is independent of the background value of $a$.\footnote{Here we have included into $\mu_0$ the wave-function renormalisation computed in appendix A. Thus, $\mu_0$ should be regarded here as a running parameter.} The second one, parametrized by the coefficient $c_\mu$, originates from the function $U$ in \eq{lagg} and scales linearly with $a$. Also $B_\mu$ is given by the sum of two contributions: one is proportional to $\mu$, with a coefficient that scales linearly with $a$; the other one scales quadratically with $a$ and comes from the function $U$.

Note that the simultaneous presence of $U$ and $\mu_0$ in the Lagrangian in \eq{lagg} breaks a continuous $R$-symmetry. By imposing such a symmetry, one could forbid $U$, forcing the coefficients $c_\mu$ and $c_B$ to vanish. However, the $R$-symmetry is explicitly broken by the parameter $m$ in \eq{polpot} and, as a consequence, by the background value of $F$.

\section{Relaxation of supersymmetry breaking}
\label{secrelax}
 
During its dynamical evolution, the relaxion scans the supersymmetry breaking scale $F\approx ma$. We are interested in the situation in which this evolution triggers a non-vanishing Higgs vacuum expectation value. The order parameter for electroweak symmetry breaking is the determinant of the Higgs mass matrix, defined as
\beq
{\mathscr D}(a) \equiv \left( m_{H_u}^2 + |\mu|^2\right)  \left( m_{H_d}^2 + |\mu|^2\right) - |B_\mu |^2  \, .
\label{mscr}
\eeq
The dependence on $a$ is contained in the soft terms, as described in \eq{mubmu} for $\mu$, $B_\mu$, while we take $m_{H_{u,d}}^2=c_{u,d} \, m^2a^2$. The coefficients $c_i$ are model-dependent, but are expected to be of order unity (unless scalar masses come from a mediation scale $M_*$ larger than $f$). 
As soft terms are running parameters, the coefficients $c_i$ have a logarithmic dependence on $a$. The expressions derived in appendix A for the soft terms should be viewed as the matching condition at $f$, the energy scale at which heavy modes are integrated out. Below $f$, the soft terms run according to the usual renormalisation-group equations and receive corrections of order $(\alpha / 4\pi )\ln f/(ma)$, where $\alpha$ refers to a generic coupling constant. In our calculation, presented in appendix B, we have neglected this logarithmic dependence. In practice, this is a conservative assumption, since the logarithmic running makes it only easier to achieve symmetry breaking by dynamical evolution. We will also assume that the coefficients $c_i$ of the operators responsible for squark and slepton masses remain positive throughout the evolution, such that colour or electric charge breaking vacua are avoided.

We consider an initial condition for the relaxion such that $a\gg \mu_0 /m$. In this situation, $\mu_0$ can be neglected in \eq{mscr} and we find the simple scaling ${\mathscr D}(a) \propto a^4$. We require the proportionality factor to be positive, since we want that electroweak symmetry is initially unbroken. The value of $a$ will progressively decrease, as the relaxion evolves. Once $a$ approaches $\mu_0/m$, the $a^4$-scaling is violated and, under certain conditions on the coefficients $c_i$, the order parameter ${\mathscr D}(a)$ can flip sign and trigger electroweak breaking. We call $a_*$ the value of the relaxion field for which ${\mathscr D}(a_*)=0$. Parametrically, it is given by
\beq
a_*=\frac{\mu_0}{m}\, c_* \, ,
\label{astar}
\eeq
where $c_*$ is a coefficient of order unity.
In appendix B we show the expression of $c_*$ in terms of the soft-terms coefficients $c_i$, together with the conditions necessary to have solutions for $a_*$. The calculation demonstrates that electroweak breaking can be consistently achieved in a fairly broad range of parameters.

In the proximity of $a_*$, the Higgs potential for the real scalar components of the two Higgs doublets can be written as
\beq
V=\frac{m_h^2}{2}\, h^2+\frac{\lambda}{4}\, h^4+\frac{m_H^2}{2}\, H^2 + \hbox{$H$-interactions} \, ,
\eeq
\beq
m_h^2=\frac{{\mathscr D}(a)}{m_H^2}\, , ~~~\lambda =\frac{g^2+g'^2}{8}\cos^2 2\beta
\, , ~~~m_H^2 = m_{H_u}^2 +  m_{H_d}^2 + 2|\mu|^2 \, ,
\label{defmh}
\eeq
where we have kept only the leading-order terms in ${\mathscr D}$.
Here $h$ and $H$ are the two mass eigenstates obtained from the current eigenstates by rotation of an angle $\beta$, with $\tan^2\beta = m_{H_d}^2/m_{H_u}^2+\ord ({\mathscr D}/m_H^4)$. For a separation of scales $|{\mathscr D}|\ll m_H^4 $, the heavy Higgs $H$ decouples and $h$ behaves like the SM Higgs. For negative ${\mathscr D}$, the SM Higgs gets a vacuum expectation value $\langle h^2\rangle = - {\mathscr D}/(\lambda m_H^2)$.

After electroweak breaking, QCD instanton effects generate a potential for the relaxion that respects only a discrete shift symmetry for $a$. Adding this interaction term to the PQ breaking potential in \eq{che}, we obtain
\beq
V(a)=\frac{m^2f^2}{2}\, a^2 +\Lambda^4 \cos a \, ,
\label{pinpot}
\eeq
where $\Lambda$ is the typical scale emerging from non-perturbative effects. An important observation of ref.~\cite{GKR} is that $\Lambda^4$ scales roughly linearly with the Higgs vacuum expectation value. Therefore, as $a$ evolves below $a_*$, the first term in \eq{pinpot} decreases, while the second one quickly increases because $|{\mathscr D}|$ is growing. A local minimum of the relaxion potential is generated when the barrier heights (measured by $\Lambda^4$) have grown enough to make sure that the two terms in $V'(a)$ can cancel each other. This happens when $\Lambda^4$ has reached the size $f^2m^2a_*$ so that $V'(a)=0$:
\beq
m \approx  \frac{\Lambda^4}{f^2\, \mu_0}      ~~~~~~~~~\hbox{(local minimum}) \, .
\label{pif1}
\eeq

So far, our discussion has been purely classical. However, at the first local minimum, the barrier height is sufficiently small to make quantum fluctuations important. Once the relaxion has established itself in its final minimum, quantum tunnelling is not a problem. We estimate that the probability of vacuum decay during the past light-cone of the observable universe is $P\sim \tau_U^4 f^4 \exp (-f^4/\Lambda^4)$, where $\tau_U$ is the present lifetime of the universe. This is completely negligible, as a result of the considerable field distance between two consecutive vacua (of size $2\pi f$) with respect to the typical available energy ($\Lambda^4$). More problematic are the quantum fluctuations at the time the relaxion is settling into its vacuum. Quantum evolution appears to populate different minima. Although all of them have roughly the same value of the weak scale, this could result in a universe made of patches with different Higgs values. This potential cosmological problem is generic of the relaxation mechanism and is present also in the original model of ref.~\cite{GKR}.

There is an important difference in our setup with respect to the non-supersymmetric case. In the model of ref.~\cite{GKR}, the barrier heights grow as we probe smaller field values and never cease to exist. As shown in appendix B, this is not the case for the relaxation of the supersymmetry scale. Quite generically, ${\mathscr D}(a)$ flips sign again at a value $a=a_{**}$ (with $a_{**}<a_*$) and turns back positive, restoring electroweak symmetry. This is consistent with the notion that a supersymmetric vacuum exists at $a=0$. As a result, for the relaxation mechanism to work, the barriers have to grow sufficiently high during the evolution between $a_*$ and $a_{**}$ and stop the relaxion before it could slide into the region with $\langle h \rangle =0$ and $a<a_{**}$. Let us investigate the issue.

Between two adjacent troughs of the periodic potential, the relaxion changes by an amount $\Delta a =2\pi$ and the soft mass scale by $\Delta F\approx m\Delta a$, a very small variation. In the vicinity of $a_*$, the Higgs mass $m_h^2 \sim {\mathscr D}/\mu_0^2$ changes more rapidly, $\Delta m_h^2\sim m\mu_0\, \Delta a$ because, while ${\mathscr D}$ happens to be near a zero, its derivative is generic. As the relaxion scans the range between $a_*$ and $a_{**}$, it crosses a huge number of oscillations $N=(a_* -a_{**})/2\pi \sim \mu_0 / m$. In doing so, the first term in the potential in \eq{pinpot} has only a modest relative variation, while the second one grows fast, since the Higgs vacuum expectation value changes from 0 to about $\mu_0$. So there is enough freedom to choose a suitable value of $m$ such that the barrier heights $\Lambda^4$ have the chance to grow up to values of order $f^2m\mu_0$ before the relaxion completes $\mu_0 / m$ periods of oscillation.

\section{Inflationary dynamics}
\label{secinf}
The dynamics of inflation is an essential element of the relaxation mechanism because it provides the friction term that stops the relaxion at a local minimum. However, the inflationary sector is not included in the Lagrangian in \eq{lagg} and here is only treated as a spectator that provides a nearly constant Hubble rate $H$. Nonetheless, the value of $H$ is subjected to several strong constraints, which limit the allowed range of the parameters in the theory. We list here these constraints, which must be satisfied for values of the relaxion field in the range $a\approx a_*$, with $a_* \approx \mu_0/m$.

The first requirement is that the relaxion satisfies the slow-roll condition during evolution. Since the slow-roll parameters are $\epsilon =\eta =2M_P^2/(a^2f^2)$, we find
\beq
m< \frac{\mu_0 \, f}{M_P}~~~~~~~~~\hbox{(relaxion slow roll)} \, .
\label{pif2}
\eeq

The requirement that the relaxion potential energy ($m^2f^2a^2/2$) is subdominant with respect to the inflaton energy ($3H^2M_P^2$) implies
\beq
H\,  > \frac{\mu_0 \, f}{M_P} ~~~~~~~~~\hbox{(inflaton dominates the vacuum energy)} \, .
\label{pif3}
\eeq

The vacuum energy that drives inflation breaks supersymmetry and can be described by the auxiliary component $F_I=HM_P$ of a chiral superfield $I$ containing the inflaton as scalar component. This source of supersymmetry breaking will feed into the soft terms of the SM fields through interactions that cannot be weaker than gravity, thus giving
\beq
\frac{1}{M_P^2}
\int d^4 \theta \, I^\dagger I \, \Phi_i^\dagger \Phi_i = H^2 \phi_i^*\phi_i \, .
\label{spit}
\eeq
In order not to spoil the relaxation of the supersymmetry scale, the soft masses in \eq{spit} must be subleading with respect to the contribution from the relaxion superfield. This implies
\beq
H < \mu_0   ~~~~~~~~~\hbox{(soft terms from inflaton are subleading)} \, .
\label{pif4}
\eeq

We require that the Hubble rate be smaller than the QCD scale $\Lambda$ to insure the formation of the potential barriers from instanton effects
\beq
H < \Lambda  ~~~~~~~~~\hbox{(potential barriers from QCD)} \, .
\label{pif5}
\eeq
This condition implies that \eq{pif4} is automatically satisfied, as $\mu_0>\Lambda$.

Finally we impose that the evolution of the relaxion is governed by the classical potential, rather than following a random walk driven by quantum fluctuations. This requires that the classical force ($V'(a)/f=m^2fa$) dominates over the stochastic term ($3H^3/2\pi$) in the relaxion equation of motion. This implies
\beq
H^3 <  m\, f \, \mu_0  ~~~~~~~~~\hbox{(classical evolution)} \, .
\label{pif6}
\eeq

We can now put together eqs.~(\ref{pif1})--(\ref{pif6}) and identify the acceptable range of the theory parameters.  The PQ scale can vary in the range $f\sim 10^9$--$10^{12}$~GeV, the so-called axion window, satisfying present experimental and cosmological bounds (for reviews see ref.~\cite{axionrev}). The value of the PQ-breaking mass $m$ is derived from \eq{pif1}
\beq
m\approx \left( \frac{\Lambda}{300~\mev}\right)^4\left( \frac{10^9~\gev}{f}\right)^2\left( \frac{10^5~\gev}{\mu_0}\right)\, 10^{-25}~\gev \, .
\label{massina}
\eeq
Taking together eqs.~(\ref{pif3}) and (\ref{pif5}), we find
\beq
H < 300~\mev ~~~~~~~{\rm and}~~~~~~~\mu_0<\left( \frac{10^9~\gev}{f}\right)\, 10^9~\gev \, .
\label{weaker}
\eeq
A stronger constraint is obtained from \eq{pif6}
\beq
H < \left( \frac{\Lambda}{300~\mev}\right)^{4/3}\left( \frac{10^9~\gev}{f}\right)^{1/3}\, 0.2~\mev ~~~~~~~{\rm and}
\nonumber
\eeq
\beq
\mu_0<\left( \frac{\Lambda}{300~\mev}\right)^{4/3} \left( \frac{10^9~\gev}{f}\right)^{4/3}\, 5\times 10^5~\gev \, .
\label{stronger}
\eeq
The stronger constraint in \eq{stronger} comes from the requirement of a classical relaxion evolution. Although justified, this condition may be too restrictive and the relaxation mechanism may operate also in presence of sizeable quantum fluctuations. For this reason, we have quoted separately the two bounds in eqs.~(\ref{weaker}) and (\ref{stronger}). Finally, note that eqs.~(\ref{pif2}) and (\ref{pif4}) do not add any information, since they are automatically satisfied when the other conditions are met.  

A successful relaxation mechanism requires that the relaxion scans a range $\Delta a$ larger than $a_*$. This implies
\beq
\Delta a > a_* =\left( \frac{300~\mev}{\Lambda}\right)^4\left( \frac{f}{10^9~\gev}\right)^2 \left( \frac{\mu_0}{10^5~\gev}\right)^2  \, 10^{30}\, ,
\label{su}
\eeq
which corresponds to an excursion $\Delta a\, f$ of $10^{39}$~GeV, for the same reference values of \eq{su}. The number of e-folds required for this field excursion, in the slow-roll regime, is given by $N=3H^2f^2\Delta a/V'(a)$. Using the lower bound on $H$ from \eq{pif3} and expressing $m$ through \eq{pif1}, we find
\beq
N > \left( \frac{300~\mev}{\Lambda}\right)^8\left( \frac{f}{10^9~\gev}\right)^6 \left( \frac{\mu_0}{10^5~\gev}\right)^4 \, 10^{42} \, .
\eeq
This enormous number of e-folds is a consequence of the shallowness of the relaxion potential caused by the tiny value of its mass $m$, see \eq{massina}.

\section{Remarks on the UV completion}
\label{secUV}

\subsection{Planckian effects}
The vastly super-Planckian field excursion required by the relaxation mechanism, see \eq{su}, casts doubts on the use of the effective field theory, since the Lagrangian in \eq{lagg} is valid only up to energies of order $f$. More generally, as the relaxion explores the super-Planckian regime, the quantum field theory approach may seem questionable. However, if we require that the potential energy does not exceed the cutoff scale ($V< f^4$), we obtain that field excursions up to $a< f/m$ are allowed. The relaxation mechanism requires $a\approx \mu_0/m$, and so the condition is satisfied. For the same reason, we can argue that a description of the relaxion evolution in the context of quantum field theory suffices and no knowledge of quantum gravity is needed, as the typical potential energy ($V^{1/4} \sim (\mu_0 f)^{1/2}$) is much smaller than the Planck mass.

Nevertheless, these considerations are not sufficient to believe that super-Planckian physics is not going to modify the relaxion potential. A first concern is that gravity is expected to violate global symmetries~\cite{globalnum} and can lead to Planck-suppressed operators that do not respect the shift symmetry, giving enhanced contributions to the potential in the super-Planckian domain.
From a low-energy point of view, this problem may be circumvented. To see this, let us assume that a small parameter $\epsilon$ controls the breaking of the shift symmetry and, because of a selection rule, the relaxion field $a$ always appears multiplied by $\epsilon$ in the potential. Using dimensional analysis, our assumption states that $V(a)/f^4=h_V(\epsilon a)$, where $h_V$ a generic function. In the specific case of \eq{polpot}, the small parameter is $\epsilon = m/f$. If gravity respects the selection rule, Planckian operators are expected to modify the potential in the form $V(a)/f^4=h_V(\epsilon a)[1+(\epsilon a f/M_P)^n]$, for any power $n$. In the case we have considered in our paper, the typical value of the expansion parameter is $\epsilon a f/M_P \sim \mu_0/M_P$, which is much smaller than one. Thus, a selection rule could keep Planckian corrections under control. Nevertheless, we will show later that the selection rule may be violated in realistic examples. 

Another concern is related to the conjecture of gravity as the weakest force~\cite{weakest}. This conjecture can be stated as follows. If a gauge force with coupling $g$ is present at low energy, then the effective field theory ceases to be valid at energies around $E\sim gM_P$. This occurs because the effective theory becomes inconsistent with gravity unless new states are added at the cutoff scale. The conjecture may be extended to non-gauge forces, in particular prohibiting super-Planckian displacements of axion-like particles, since their decay constants $f$ are ruled to be smaller than $M_P$~\cite{weakestax}. In our case, the super-Planckian excursion is not due to $f>M_P$, but to the relaxion monodromy. The violation of gravity as the weakest force comes from the very small parameter $\epsilon = m/f$ characterising the strength (or, more precisely, the weakness) of the force breaking the shift symmetry. The smallness of $\epsilon$ is ultimately related to the super-Planckian displacement of the relaxion. In summary, the conjecture of gravity as the weakest force indicates a premature violation of the effective theory at a scale $\epsilon M_P$. Therefore, the mechanism of relaxation is incompatible with the conjecture of gravity as the weakest force.

\subsection{Generalising the relaxion potential} 
On the positive side, the relaxation mechanism is rather robust in the sense that it does not require a very special structure for the relaxion potential. Even if above $M_P$ the form of the potential is not the same as the one we assumed at lower energy, the mechanism can operate nonetheless. To explain the point, let us reconsider our analysis in terms of $\epsilon$ and assume that the expression of the soft masses satisfies the same selection rule. Dimensional analysis in the effective theory then gives
\beq
V=f^4\, h_V (\epsilon a) \, , ~~~~~{\tilde m}=  f\, h_{\tilde m} (\epsilon a) \, ,
\label{general}
\eeq
where $h_V$ and $h_{\tilde m}$ are generic functions. For concreteness, we take $h_V(x)=x^n$ and $h_{\tilde m}(x)=x^p$. 
We can now repeat the analysis of section~\ref{secrelax} and \ref{secinf} for the more general case defined in \eq{general}. We will recover our previous results for $n=2$, $p=1$, and $\epsilon = m/f$.

The critical value for $a$ at which electroweak symmetry is first broken is determined by the condition ${\tilde m} (a_*)\approx \mu_0$, which gives
\beq
a_* \approx \frac{1}{\epsilon} \left( \frac{\mu_0}{f}\right)^{\frac{1}{p}}
    ~~~~~~~~~\hbox{(critical condition for EW breaking)} \, .
\label{pifg0}
\eeq
The local minimum of the relaxion potential is reached for
\beq
\epsilon \approx  \left(\frac{\Lambda}{f}\right)^4 \left( \frac{f}{\mu_0}\right)^{\frac{n-1}{p}}      ~~~~~~~~~\hbox{(local minimum}) \, .
\label{pifg1}
\eeq
The constraint that the relaxion potential is smaller than the inflaton energy gives
\beq
H\,  > \frac{f^2}{M_P} \left( \frac{\mu_0}{f} \right)^{\frac{n}{2p}}~~~~~~~~~\hbox{(inflaton dominates the vacuum energy)} \, .
\label{pifg3}
\eeq
Classical evolution of the relaxion at early times requires
\beq
H <  \frac{\Lambda^{4/3}}{f^{1/3}}  ~~~~~~~~~\hbox{(classical evolution)} \, .
\label{pifg6}
\eeq
Combining eqs.~(\ref{pifg3}) and (\ref{pifg6}), we find the upper limit on $\mu_0$
\beq
\mu_0 < \frac{\Lambda^{\frac{8p}{3n}}\, M_P^{\frac{2p}{n}}}  {f^{\frac{14p}{3n}-1}} \, .
\label{supbound}
\eeq
Supersymmetry implies $n=2p$. In this case, the bound on $\mu_0$ is independent of the special value of $n$ and the result in \eq{supbound} coincides with \eq{stronger}. This means that the relaxation of supersymmetry breaking works independently of the form of the superpotential that breaks the shift symmetry.

\subsection{Supergravity effects}
Our conclusions about the robustness of the mechanism come from estimates based on the effective theory. However, a UV completion that introduces new dimensionful couplings (like gravity) and violates the selection rule can modify our conclusions. To illustrate the point, take the example of supergravity, in which the scalar potential is given by
\beq
V=\frac{e^{\frac{f^2}{M_P^2}\, K}}{f^2} \left( {K''}^{-1} \left| W' 
+ \frac{f^2}{M_P^2}K' \, W\right|^2 -\frac{3f^2}{M_P^2}\, |W|^2 \right) \, ,
\label{sugra}
\eeq
where $K$ is the (dimensionless) K\"ahler potential defined in \eq{lagg} and $W$ is the superpotential. 
Since we have assumed that the breaking of the shift symmetry resides only in the superpotential, $K$ is a function of $s$ alone, while $W$ depends also on $a$. 

Suppose, as done previously, that $W$ is a function of $\epsilon a$, where $\epsilon$ measures the breaking of the shift symmetry. In this case $W'$ is typically suppressed by a factor of $\epsilon$ with respect to $W$ and we obtain
\beq
V(a)\approx -\left( 3-\frac{f^2\, {K'}^2}{M_P^2\, K''}\right)  \frac{|W|^2}{M_P^2} \, .
\label{sugraspit2}
\eeq
If $\langle s\rangle =\ord ( 1)$, then ${K'}^2/K'' =\ord ( 1)$ and the potential develops an unstable direction as $a$ grows. This feature is well known in supergravity inflationary models and it usually requires the addition of new stabiliser fields~\cite{sugrainf}. In our case, this runaway direction is a virtue because it could naturally explain the initial condition of the relaxion in the early universe. Assuming that the relaxion starts at Planckian values, the potential in \eq{sugraspit2} would make it slide along the runaway direction deep into the super-Planckian region until it is stopped by QCD effects. The huge value of $a$ needed by the relaxation mechanism would not be the result of an artificial choice of initial conditions, but would be derived from the dynamical evolution.

Unfortunately, supergravity brings in a problem that was not manifest in the effective theory analysis. Since the soft masses are $\tilde m \sim |W|/M_P^2$ and the potential is $V\sim |W|^2/M_P^2$, we obtain $V\sim \tilde m^2 M_P^2$, independently of the specific form of the superpotential $W$. The constraint that the energy density is inflaton-dominated then requires $H>\tilde m$, preventing the relaxation mechanism. Of course, in order to understand if this is a serious impediment to relaxation in supergravity, one should have control over the mechanism that cancels the cosmological constant. Nevertheless, we believe that this example illustrates the difficulties that one could encounter when violations of the selection rule modify the expectations based on the effective theory. In the case of supergravity, this comes about because $W$ and $W'$ (where $W'=dW/da$) cannot both depend on the single variable $\epsilon a$.

\subsection{Effects beyond quantum field theory}

There is another important issue about the UV completion of the theory we want to remark upon. The effective theory defined by the Lagrangian in \eq{lagg} has a validity cutoff at the relatively low scale $f$, where the relaxion interactions lead to a violation of perturbative unitarity and require a UV completion. However, we are assuming monodromy for the relaxion and it is not clear if a consistent UV completion within quantum field theory exists. One could then believe that our theory requires a drastic departure from quantum field theory, maybe involving string theory, at the scale $f$. We want to claim that this is not the case.

Let us first consider the theory in the limit $\epsilon \to 0$, shutting off the effects of monodromy.  In this case we simply recover the usual axion interactions.  This theory must be UV-completed at the scale $f$, otherwise the derivative couplings of the relaxion would lead to parametric growth of scattering amplitudes at energies above $f$.  We know how to UV-complete the axion-like interactions by promoting the theory to a renormalisable model which spontaneously breaks a global PQ-symmetry.  In this case the relaxion will typically be identified as the argument of a complex scalar field and the additional heavy radial mode will enter scattering amplitudes, curing any pathological behaviour in physical processes.  With this in mind, although in this work we always consider only the light degrees of freedom remaining in the low-energy effective theory, we envisage such a UV-completion for the axion-like couplings to enter at the scale $f$.  In our case, this would be a supersymmetric axion model.

We now introduce the shift-symmetry breaking terms controlled by the small parameter $\epsilon$.  Let us consider a theory where the interactions leading to non-compact behaviour are present, while the ordinary axion-like interactions are UV-completed at the scale $f$ as described above.  
As the only two parameters in this simplified picture are $f$ and $\epsilon$, we expect any pathological behaviour of scattering amplitudes to become apparent only at a scale
$\Lambda_{PQ} \sim ({f}/{\epsilon})^k f$,
where $k$ is some power characteristic of a given process.  In the limit $\epsilon \to 0$, we recover the result that, leaving gravity aside, the theory is UV complete up to arbitrarily high energies. Since $\epsilon$ is very small, $\Lambda_{PQ}$ is so large that, although the relaxion potential is exotic from a field theory perspective, we do not expect it to exhibit pathological behaviour until well above the Planck scale, at which point field theory may break down in any case.

In conclusion, although a UV-completion of the relaxion monodromy would provide valuable insight and understanding of the possible UV physics behind the mechanism, such a UV-completion is not urgently required for the application of the relaxion mechanism at energies below the Planck scale.

\section{The structure of supersymmetry breaking}
\label{secstru}
To understand the supersymmetry breaking mechanism it is necessary to study the vacuum structure of the theory.  The process is complicated in comparison to more familiar tree-level supersymmetry breaking scenarios by the fact that the dynamics stabilising the relaxion in a metastable minimum arises at the QCD scale, whereas the soft masses are induced at much higher energy.  Very different energy scales must be considered simultaneously, since all of them play a role in the mechanism.  Furthermore, the stabilisation of the relaxion in a supersymmetry-breaking minimum depends on non-perturbative instanton effects which may only be estimated from the chiral Lagrangian.

As all of the relevant relaxation dynamics occurs at energy below the QCD scale, we wish to determine the potential for the lightest degrees of freedom: the relaxion and the relaxino.  To begin with we will construct a supersymmetric effective theory below the weak scale, including light quarks and the relaxion multiplet. At this stage, supersymmetry is essential to retain the correct properties of the relaxion and relaxino couplings. Then we will integrate out the heavy squarks to determine the effective theory at the QCD scale.  Finally, to go below the QCD scale we will include the chiral condensate and study its effects on the relaxion and relaxino.

Our starting effective theory below the weak scale includes the relaxion multiplet and light quarks (together with gluons and photons, although we do not write them explicitly). The Higgs multiplets have been integrated out and the effect of chiral symmetry breaking from the Higgs vacuum expectation value is captured through the quark mass $m_q$. In order to keep track of the relations between the couplings of the relaxion and the relaxino, we will make supersymmetry manifest and describe the theory in terms of the chiral superfields $S$ (relaxion), $Q$ (quark) and $Q^c$ (anti-quark). To illustrate the mechanism more clearly, we will not use the most general K\"ahler potential, as done in \eq{lagg}, but retain only the essential terms, which are
\be
K = \frac{f^2}{2} (S + S^\dagger)^2 + Q^\dagger Q + Q^{c \dagger} Q^c- \frac{f^2 (S + S^\dagger)^2}{{M_*}^2} ( Q^\dagger Q +Q^{c \dagger} Q^c) - \frac{f^2 (S + S^\dagger)^4}{4!} \, .
\label{eq:kahltoy}
\ee
In a basis in which the relaxion couplings have been rotated into the quark mass terms, the superpotential is given by the sum of a PQ-invariant quark mass term and the PQ-breaking relaxion term in \eq{polpot}
\be
W = m_q e^S Q^c Q + \frac{m}{2} f^2 S^2 \, .
\label{eq:superpottoy}
\ee

The background value of the relaxion gives a non-vanishing value to the auxiliary field $F=ima/\sqrt{2}$. As a result, supersymmetry-breaking terms are induced. Squarks and the scalar srelaxion $s$ acquire the soft masses\footnote{We recall that we are taking $S$ to be dimensionless. This choice makes the dimensions of the interactions look unusual. Ordinary dimensions are recovered by recalling that the physical fields are $fa$, $fs$, and $f\tilde a$. Moreover, we use two-component Weyl spinors for fermions.} 
\be
\Lag \, \supset \, 
-\frac{a^2 f^2 m^2}{M_\ast^2} \left( |\tilde{q}|^2 + |\tilde{q}^c|^2 \right)  
-\frac{a^2 f^2 m^2}{2} s^2
\, ,
\label{squamas}
\ee
together with a chirality-flipping squark mass term
\be
\Lag \, \supset\, \frac{i m_q }{\sqrt{2}}\, a m\, e^{i a/\sqrt{2}}\,  \tilde{q}^c \tilde{q} + \hc
\label{eq:squarkmix}
\ee
Although the srelaxion $s$ is a singlet scalar, no tadpole term is induced.  
The relevant chirality-preserving (but supersymmetry breaking) and chirality-flipping (but supersymmetry preserving) Yukawa interactions between relaxino, quarks, and  squarks are
\be
\Lag \, \supset \, -\frac{i \sqrt{2}}{M_\ast^2} a m f^2 \left( \tilde{q}^\ast \, \tilde{a}  q + \tilde{q}^{c \ast}\, \tilde{a} q^c \right) - m_q e^{i a/\sqrt{2}} \left( \tilde{q}^c \, \tilde{a}  q + \tilde{q}\, \tilde{a}  q^c \right) +\hc ~~,
\label{eq:squarkint}
\ee
and the Yukawa couplings of the srelaxion to quarks and the relaxino are
\be
\Lag \, \supset \, -  \frac{m_q}{\sqrt{2}} \, e^{i a/\sqrt{2}} s\, q^c q -\frac{i}{2} a m f^2 s\, \tilde{a}\tilde a  +\hc 
\label{eq:sint}
\ee

\begin{figure}[t]
\centering
\includegraphics[height=2.9in]{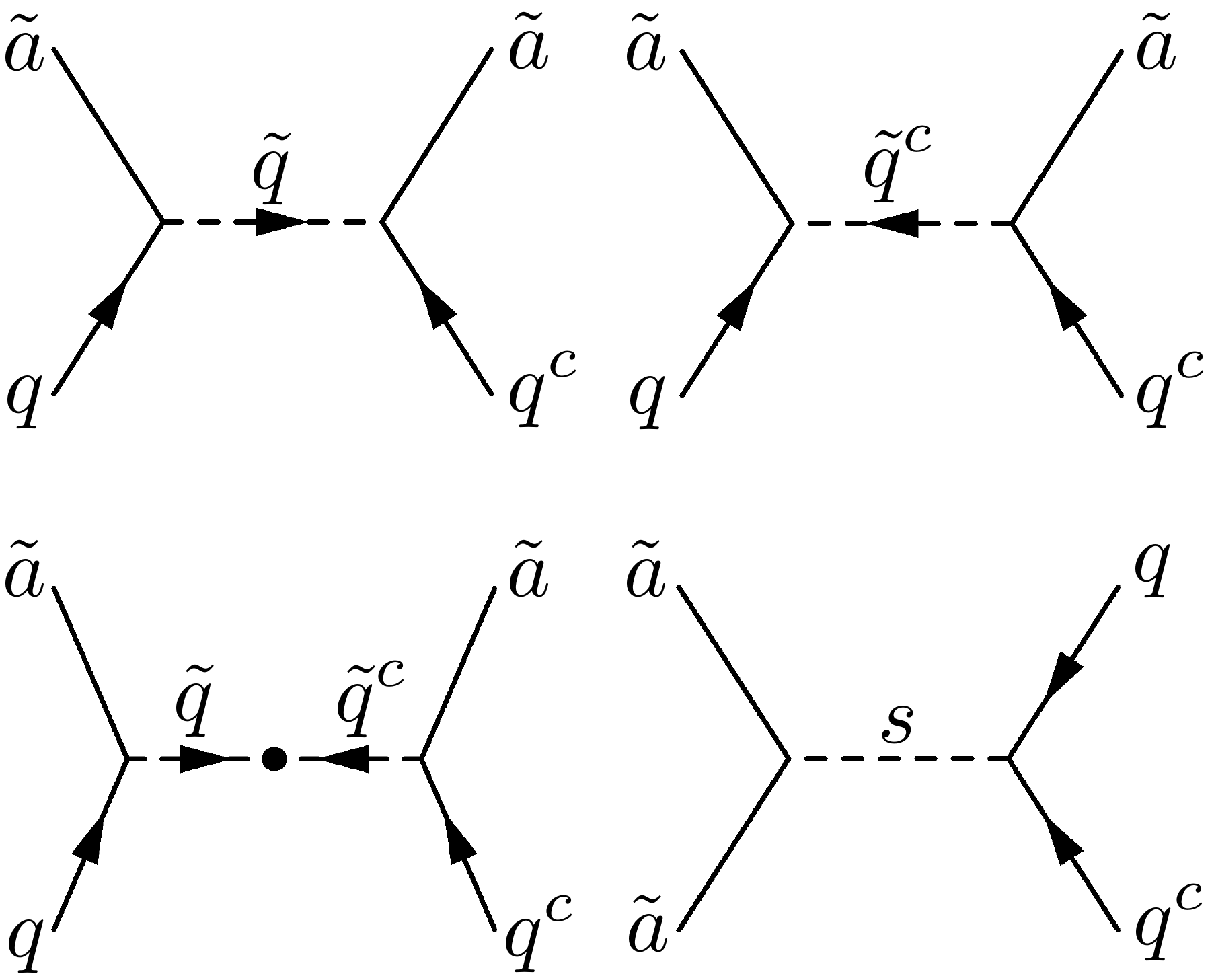}
\caption{The Feynman diagrams that generate the four-fermion operator $(q^cq)\, (\tilde{a} \tilde{a})$, after integrating out the squarks and the srelaxion $s$.}
\label{fig:bodydiag}
\end{figure}

By construction the soft masses are well above the weak scale, thus to understand the theory at the QCD scale the squarks and the srelaxion must be integrated out.  Due to the interactions of eqs.~(\ref{eq:squarkint})--(\ref{eq:sint}), squark and srelaxion exchange generates four-fermion interactions involving the relaxino and quarks, as described by the Feynman diagrams in \fig{fig:bodydiag}. Working at the leading order in $m_q$, we can treat the mixing term in \eq{eq:squarkmix} as a mass insertion. The dependence on the messenger mass $M_\ast$ drops out from the coefficients of the four-fermion interactions. In the two diagrams with chirality-preserving squark propagators, the $1/M_\ast^2$ factor appearing in the chirality-preserving Yukawa coupling is cancelled by the squark mass. In the diagram
with mass insertion, the factor $1/M_\ast^4$ from the two Yukawa vertices is cancelled by the double squark propagator.  The dependence on the srelaxion soft mass also cancels in the same way.

After summing the contributions from squark and srelaxion exchange and accounting for the Fierz identity $(\tilde{a} \tilde{a}) \, ( q^c q) = -2 (\tilde{a} q) \, (\tilde{a} q^c)$, the induced four-fermion operator is
\be
\Lag_{4f} = - \frac{1 }{2 \sqrt{2} a m}\left(   im_q e^{{i a}/{\sqrt{2}}}  \,  q^c q + \hc \right) \left( \tilde a \tilde a+ \hc \right)
\label{eq:4f}
\ee

Combining this with the PQ-breaking mass terms and the interactions of the relaxion, the low-energy theory with supersymmetric states integrated out is described by
\be
\Lag   =    -\frac{m^2}{2} f^2  a^2 -  \frac{m}{2} f^2 \left( \tilde{a}\tilde a + \hc \right) -\left(
m_q\,  e^{{i a}/{\sqrt{2}}} \,  q^cq + \hc \right) + \Lag_{4f}  \, .
\label{eq:relaxionpotcoup}
\ee

To study the theory below the QCD scale, we capture the non-perturbative QCD effects that generate the quark chiral condensate through the replacement $m_q \langle q^c q\rangle \to \Lambda^4/2$.  This leads to the following effective Lagrangian for the relaxion and relaxino\footnote{For simplicity we have not included the SM pion fields which have a small mixing with the axion. This simplification does not modify our result, which is not affected by the inclusion of the pions.} 
\be
\Lag = -V(a) -  \frac{m_{\tilde a}(a)}{2} f^2 \left( \tilde{a}\tilde a + \hc \right)
\ee
\beq
V(a)= \frac{m^2}{2} f^2  a^2 + \Lambda^4 \cos \frac{a}{\sqrt{2}} \, ,~~~~~~~~
m_{\tilde a}(a) = m-\frac{\Lambda^4 \sin \frac{a}{\sqrt{2}}}{\sqrt{2}\, a mf^2} \, .
\label{eq:relaxionpotcoupch}
\eeq
Equation~(\ref{eq:relaxionpotcoupch}) exhibits how the periodic term in the relaxion potential is generated from QCD instantons. Minimising the potential $V(a)$, we obtain the vacuum expectation value of the relaxion
\beq
\left. V' (a)\right|_{a=\langle a \rangle} =0 ~~~~~~\Rightarrow ~~~~~~
m^2 f^2 \langle a \rangle = \frac{\Lambda^4}{\sqrt{2}} \sin \frac{\langle a\rangle}{\sqrt{2}} \, .
\eeq 
On the vacuum, the effective mass of the relaxino, defined in \eq{eq:relaxionpotcoupch}, exactly vanishes since $m_{\tilde a}(\langle a\rangle ) =0$. 

This completes our consistency check, as in global supersymmetry we know that spontaneous supersymmetry breaking must be accompanied by a massless Goldstino. We conclude that supersymmetry is spontaneously broken at the metastable vacuum generated by non-perturbative QCD effects and the Goldstino can be identified with the relaxino.

\section{The relaxino (alias gravitino)}
\label{secgravitino}
During the cosmological evolution of the relaxion, its supersymmetric partner (the relaxino) remains light, with mass of order $m$. In this phase, the Goldstino resides primarily in the inflationary sector, as dictated by the condition of \eq{pif3} that the inflaton dominates the vacuum energy. As shown in section~\ref{secstru}, once the relaxion is stabilised, it plays the role of Goldstino. However, gravity insures that its degrees of freedom are absorbed in the spin-1/2 components of the gravitino, which acquires a mass
\beq
m_{3/2} = \frac{F\, f}{\sqrt{6} M_P}\, .
\eeq
Here $F\approx ma$ is the typical mass scale of supersymmetric partners. To keep track of the model-dependence of the supersymmetric mass spectrum, we define
\beq
\tilde m = k \, F \, ,
\eeq
where $\tilde m$ is the physical mass of the sparticle and $k$ can be different for each individual sparticle. For instance, we have found in section~\ref{secframe} that $k\approx \alpha/4\pi$ for gauginos, while for squarks and sleptons $k$ is of order unity if the mediation occurs at the scale $f$, or $k\approx f^2/M_*^2$ otherwise. Thus, we can write
\beq
m_{3/2} = \frac{1}{k} \left( \frac{\tilde m}{10^5~\gev}\right) \left( \frac{f}{10^9~\gev}\right) 17~\kev \, .
\label{gravimas}
\eeq
This means that, depending on the parameter choice, the gravitino (or relaxino) mass varies in the keV to GeV range.

As the gravitino is a factor $f/M_P$ lighter than the other supersymmetric particles, it is the LSP. Any other sparticle ($\tilde P$) decays into the relaxino with a width
\beq
\Gamma ({\tilde P}\to P {\tilde a}) =\frac{{\tilde m}_P^5}{48\pi\, m_{3/2}^2M_P^2}\, .
\label{decaygr}
\eeq
The decay rate is too fast to be of much consequence for cosmological or astrophysical considerations, but can play a role at high-energy colliders, as discussed in section~\ref{secph}.

More interesting for cosmological applications is the relic abundance of relaxinos. Since Goldstinos have derivative couplings, in the early universe they will be more easily produced at high temperatures. Their relic abundance will then depend on the reheating temperature $\trh$ of the thermal bath produced by inflaton decays. This brings up the concern about possible upper bounds on $\trh$ from the relaxation mechanism. Thus, we turn to discuss this issue.

If $\trh$ is larger than the typical QCD scale, at the end of inflation the barriers in the relaxion potential disappear and the field $a$ keeps on sliding down its potential. This continues for a time $H_{\rm QCD}^{-1}$, the Hubble rate at the QCD phase transition, when barriers are restored. During this time, the slow-rolling relaxion has travelled a distance $\Delta a ={\dot a}H_{\rm QCD}^{-1}=V'/(3f^2H_{\rm QCD}^2)$, with $V' \sim m\, {\tilde m}f^2$. As shown at the end of section~\ref{secrelax}, the change in the Higgs mass parameter for a variation $\Delta a$ is $\Delta m_h^2/\Delta a \sim m\, {\tilde m}$. Using $H_{\rm QCD}\sim T^2_{\rm QCD}/M_P$ with $T_{\rm QCD}\sim 1~\gev$, we find that the relative change in the Higgs mass is $\Delta m_h^2/ m_h^2 \sim \Lambda^8M_P^2/(f^4T^4_{\rm QCD}m_h^2) \sim (10^9~\gev /f)^4 \times 10^{-8}$. Since this change is insignificant, we conclude that the relaxation mechanism gives no bound on $\trh$.
 
For $\trh > {\tilde m}$, the thermal relic abundance of gravitinos gives a contribution to the energy density of the universe today~\cite{grav1} (for a recent reanalysis, see ref.~\cite{grav2})
\beq
\Omega_{3/2}h^2 = \left( \frac{\trh}{10^8~\gev}\right) \left( \frac{\mev}{m_{3/2}}\right) \left( \frac{\tilde m}{10^5~\gev}\right)^2\, 2\times 10^7 ~~~~~~~~\hbox{(for $\trh >{\tilde m}$)}\, .
\label{omegaga}
\eeq
Using \eq{gravimas}, we find that $\Omega_{3/2}$ always exceeds the measured dark matter density for $f$ in the axion window. Thus we conclude that $\trh$ must be smaller than $\tilde m$.

If $\trh < \tilde{m}$, gravitinos can still be created in the early universe by pair-production from scattering of SM particles and virtual sparticle exchange. The thermal-averaged scattering rate times velocity is~\cite{fabio}
\beq
\langle \sigma (PP\to {\tilde a}{\tilde a}) v \rangle \approx \frac{T^6}{16\pi m_{3/2}^4 M_P^4}\, .
\eeq
From this, we obtain the gravitino contribution to the universe energy density~\cite{gkrgkr}
\beq
\Omega_{3/2}h^2 \approx \left( \frac{\trh}{10^4~\gev}\right)^7 \left( \frac{\mev}{m_{3/2}}\right)^3\times 10^{-15} 
~~~~~~~~~~~~~\hbox{(for $\trh <{\tilde m}$)}\, .
\label{dadada}
\eeq
From \eq{dadada} we infer that as soon as $\trh$ is sufficiently smaller than $\tilde m$, relic gravitinos are never overabundant, thanks to the steep sensitivity of $\Omega_{3/2}$ on $\trh$. In the case of a split spectrum ({\it e.g.} when gauginos are lighter than scalars by a loop factor), $\tilde m$ should be interpreted as the mass of the lightest supersymmetric partner of SM particles ({\it e.g.} the lightest gaugino). This is because  
relic gravitinos can still be produced by scattering of the light gauginos, even in the limit of heavy scalars.   

In conclusion, we find that the reheating temperature must be smaller than the mass of the lightest SM supersymmetric partner ({\it e.g.} $\trh < M_{\tilde{g}}$), or else the gravitino energy density is too large. On the other hand, the gravitino ({\it i.e.} relaxino) can explain the dark matter density if $\trh \approx M_{\tilde{g}}$.

\section{Phenomenology}
\label{secph}

The phenomenology of the relaxion and supersymmetric particles is concerned with physics at very different energy scales and, in our setup, one may observe experimental signatures in both of these two seemingly disparate experimental frontiers.

\subsection{Relaxion detection}
The relaxion can be detected in usual axion searches (for reviews see ref.~\cite{axionrev}).
The two couplings most relevant for relaxion phenomenology are the coupling to the photon, $a F \widetilde{F}$, and the coupling to the gluon $a G \widetilde{G}$.  If the relaxion comprises some fraction of the dark matter the relaxion-photon coupling can be probed with microwave cavity experiments (also referred to as haloscopes).   In the case that the relic abundance of relaxions is negligible then relaxion helioscopes, light-through-wall experiments, and observations of astrophysical objects such as stars, compact stars, and supernovae, may be used to search for the production of relaxions via the axion-photon coupling.

The relaxion-gluon coupling is directly related to the neutron electric dipole moment.  If the relaxion comprises some of the dark matter this coupling may be probed through searches for NMR effects generated by an oscillating nEDM \cite{Graham:2013gfa} or oscillating atomic and molecular EDMs \cite{Stadnik:2013raa}.  Thus it may be possible in the future to probe both the relaxion-photon and relaxion-gluon couplings.

The relaxion couplings to gauge fields come from the super potential term $C_a(S) \, {\rm Tr} {\cal W}_a{\cal W}_a$ in \eq{lagg}. The same coupling also gives rise to gaugino masses. This means that there is a relation between the relaxion coupling to photons and gluons (a low-energy physics observable) and gaugino masses (a high-energy physics observables). From the results presented in appendix A, we derive that such relation is 
\beq
\frac{ c_{ a F \tilde F }  }{  c_{ a G \tilde G }  } = 
 \frac{  \cos^2 \theta_W M_{\tilde B} + \sin^2 \theta_W  M_{\tilde W} }{  M_{\tilde g}  }  
- \frac{\alpha_{\rm EM}}{\pi} \frac{B_\mu}{\mu \, M_{\tilde g}}f \left( \frac{\mu^2}{m_{H}^2}\right)   \, .
\label{link}
\end{equation}
Unfortunately the link between relaxion couplings and gaugino masses is polluted by the gauge-mediation-like contributions to $M_{{\tilde B},{\tilde W}}$ from Higgs-Higgsino loops. This contribution is of the same order of magnitude of the first term in \eq{link} because $B_\mu /\mu$ is expected to be one-loop larger than $M_{{\tilde B},{\tilde W}}$. Loop effects are also going to modify \eq{link}. These extra contaminations make \eq{link} not very useful for phenomenological applications, although the relation is representative of possible links between low-energy and high-energy observables.

Nonetheless, due to the connections between relaxion and gaugino physics it may be possible to extract some aspects of the soft mass structure in the event of axion discovery.  For example, in this setup if a relaxion coupling to photons were observed, it would imply that at least some contribution to the bino and wino soft masses came from the $S \, {\rm Tr} {\cal W}_a{\cal W}_a$ coupling.  On the other hand, if a relaxion coupling to gluons (through an oscillating EDM) were observed and no coupling to photons were measured, then this would imply that the dominant source of the bino and wino masses was likely due to gauge-mediated effects from the Higgs sector.

\subsection{LHC Phenomenology}

Our relaxation mechanism parametrically decouples the supersymmetry-breaking scale from the weak scale, thus naturally predicting that supersymmetric particles have large masses. Nevertheless, the cosmological constraints discussed in section~\ref{secinf} imply that supersymmetric masses cannot be arbitrarily large, but must lie below some hundreds of TeV, see \eq{stronger}. This region of masses is favourable for the prediction of the Higgs mass, since its measured value gives an upper bound on the supersymmetry scale of about $10^{10}$~GeV (for a degenerate spectrum) or $10^8$~GeV (for a split spectrum)~\cite{higgslim}. The heavy sparticles also eliminate problems with flavour-violating processes and dimension-5 proton decay operators. 

In spite of having supersymmetry broken at such a high scale, all hopes for discovery at the LHC are not lost. The key point for collider phenomenology is that gauginos are expected to be lighter than squarks and sleptons by a gauge loop factor. This result is deeply rooted in the structure of the theory, since the gauge sector communicates with the relaxion sector through the quantum anomaly, which originates at one loop. Taking \eq{pippog} for the gaugino masses and expressing the scalar masses as $\tilde m = kF$ (where $k$ parametrizes the model dependence), we obtain
\bea
M_{\tilde g} &\approx& c_3 \, \left( \frac{{\tilde m}/k}{10^5~\gev}\right)\,  700~\gev \, , 
\label{gaug3}\\
M_{\tilde W} &\approx& c_2 \, \left( \frac{{\tilde m}/k}{10^5~\gev}\right)\,  250~\gev \, , 
\label{gaug2}\\
M_{\tilde B} &\approx& c_1 \, \left( \frac{{\tilde m}/k}{10^5~\gev}\right)\,  120~\gev \, ,
\label{gaug1}
\eea 
where we have chosen a GUT normalisation of the $U(1)$ gauge coupling constant.
Here $c_i$ are the anomaly coefficients defined in \eq{defanomal}. Their values depend on the PQ completion at the scale $f$. While $c_3$ must be non-zero because the QCD anomaly is an essential element of the story, $c_2$ and $c_1$ could vanish. Nevertheless, this would not change our estimate in eqs.~(\ref{gaug2})--(\ref{gaug1}) because the contribution to electroweak gaugino masses from $\mu$ is parametrically of the same order.

Notwithstanding the model dependence inherent in eqs.~(\ref{gaug3})--(\ref{gaug1}), it is clear than gauginos could be within the reach of the LHC or, at least, of future colliders. The mass spectrum that we obtained is very similar to anomaly-mediated Mini-Split~\cite{minisplit}. So our claim is that relaxation provides a framework for a natural realisation of Mini-Split. 

There are two differences with respect to the case of anomaly-mediated Mini-Split. First, the ratios of gaugino masses are not rigidly determined by the $\beta$ functions (as in anomaly mediation) but are essentially free parameters, unless one specifies the theory at the scale $f$. Second, unlike anomaly mediation where the gravitino is heavier than gauginos, here the gravitino ({\it i.e.} relaxino) is the LSP. These two features change completely the collider phenomenology. 

At hadron colliders, gluino production is the leading discovery process. As R-parity is assumed to be conserved, all gluino decays will eventually terminate with the relaxino. However, we can envisage different situations, depending on the nature of the next-to-lightest supersymmetric particle (NLSP). 
A common feature is that the NLSP decays into the relaxino with a lifetime that can be derived from \eq{decaygr} and is given by
\be
\tau_{\rm NLSP} =  \left( \frac{m_{3/2}}{1 \text{ MeV}} \right)^2  \left( \frac{1 \text{ TeV}}{M_{\rm NLSP}} \right)^5\, 1.7 \times 10^2 \text{ meters}/c \, ,
\label{eq:disp}
\ee
where $c$ is the speed of light. For an NLSP with energy $E$,
the displacement in the decay is $\ell_{\rm NLSP} =\tau_{\rm NLSP} \sqrt{ (E/M_{\rm NLSP})^2-1}$. 
Since, as discussed in section~\ref{secgravitino}, the relaxino mass $m_{3/2}$ could in principle vary between the keV and GeV range, \eq{eq:disp} predicts that a collider-produced NLSP could travel for distances $\ell_{\rm NLSP} $ that vary between 100~microns and a journey to the moon. To describe the collider signatures we will now individually consider the three possible cases of NLSP.

\subsubsection{Gluino NLSP}
If the gluino is the NLSP, it decays into a gluon and a relaxino.
Gluino pair production at the LHC would lead to two hard gluon jets and missing energy, $pp\to \tilde{g} \tilde{g} \to g g \tilde{a} \tilde{a}$.  This signature would be striking as the lack of a cascade decay chain implies a much lower jet multiplicity than in the typical Mini-Split case where each gluino decays to a neutralino through an off-shell squark.

The gluino lifetime is given by \eq{eq:disp}. As displacements $\ell_{\rm NSLP}$ greater than $100~\mu$m can be experimentally resolved, gluino decays are likely to be displaced by an observable distance.  If the decay occurs within the detector, the signature would thus be $jj+\text{MET}$, where the jet vertices are displaced.  If the decay occurs outside the detector, the gluino would appear as a long-lived coloured particle and would show up in dedicated R-hadron searches~(for a review, see ref.~\cite{rhad}).

\subsubsection{Bino NLSP}
If the bino is the NLSP, the gluino decays predominantly through an off-shell squark $\tilde{g} \to q \overline{q} \tilde{B}$ with a lifetime~\cite{lifet}
\be
\tau_{{\tilde{g} \to q \bar{q} \tilde{B}}} \approx  \left( \frac{\tilde m}{10^5~\gev} \right)^4  \left( \frac{1 \text{ TeV}} {M_{\tilde{g}}} \right)^5 ~10^{-1}~\mu{\rm m}/c\, .
\label{eq:decaybino}
\ee
For gluinos in the TeV mass range and squarks a loop factor or less above the gluino mass, it is unlikely that these decays would be observably displaced, unless $c_3$ in \eq{gaug3} is sufficiently small.  The bino decays via $\tilde{B} \to \gamma \tilde{a}$ or $\tilde{B} \to Z \tilde{a}$ and the displacement for this decay is also determined by \eq{eq:disp}, thus it is observable. The decay $\tilde{B} \to h \tilde{a}$ is highly suppressed by the negligible bino-Higgsino mixing.  Because Higgsinos are heavy and gauginos are almost pure states, there is a prediction for the branching ratio of the bino decays
\be
\frac{\Gamma (\tilde B \to Z \tilde a )}{\Gamma (\tilde B \to \gamma \tilde a )}=\tan^2\theta_W \left( 1-\frac{m_Z^2}{m_{\tilde B}^2} \right)^4 ~~,
\ee
which, if measured, could confirm the bino nature of the NLSP.

In summary, for a bino NLSP the collider signature would be striking: $jjjj+\gamma \gamma + \text{MET}$ where the photons would emerge from a displaced vertex. This signal can also occur in models with gauge mediation.  If kinematically available, one or both of the photons may be replaced by a Z-boson.  In addition, if the squarks were heavy enough the jet pairs may also be displaced, leading to a final state with four displaced vertices.

\subsubsection{Wino NLSP}
If the wino is the NLSP, the gluino decays into a pair of quarks and a charged or neutral wino with a lifetime given by \eq{eq:decaybino}.  Due to the absence of mixing with the Higgsinos, the charged and neutral mass splitting is $M_{\tilde{W^\pm}}-M_{\tilde{W^0}}\approx 165$ MeV at two loops~\cite{Ibe:2012sx}.  The consequence is that, if the charged wino is produced in a cascade, it will decay via $\tilde{W}^\pm \to \pi^\pm + \tilde{W}^0$ where the pions are too soft to be reconstructed at the LHC.  This decay would occur with a lifetime $\tau_{{\tilde{W}^\pm \to \pi^\pm + \tilde{W}^0}} =6$~cm$/c$ \cite{Ibe:2012sx}.  This leads to a very interesting scenario as the typical displacement for charged ($\tilde{W}^\pm \to {W}^\pm + \tilde{a}$) or neutral ($\tilde{W}^0 \to \gamma/Z + \tilde{a}$) wino decay to the relaxino is given by \eq{eq:disp} and for the charged wino the branching ratio for this decay may exceed or fall short of the decay to pions, leading to a number of distinct signatures.
\begin{itemize}
\item  Both gluinos decay to neutral winos.  This would look similar to the bino NLSP and the collider signature would again be: $jjjj+\gamma/Z \,  \gamma/Z + \text{MET}$ where the $\gamma/Z$ could be displaced.   Again, if the squarks were heavy enough the jet pairs may also be displaced, leading to a final state with four displaced vertices.  Although the collider topologies are similar, the ${\tilde W}^0$ NLSP could be distinguished from the $\tilde B$ NLSP scenario through the branching ratio prediction
\be
\frac{\Gamma (\tilde W^0 \to Z \tilde a )}{\Gamma (\tilde W^0 \to \gamma \tilde a )}=\cot^{2}\theta_W \left( 1-\frac{m_Z^2}{m_{\tilde W}^2} \right)^4 \, .
\ee
\item  One or both gluinos decay to charged winos and $\ell_{\tilde{W}^\pm \to \pi^\pm + \tilde{W}^0} \ll \ell_{\tilde{W}^\pm \to {W}^\pm + \tilde{a}}$.  In this regime the majority of charged winos produced from gluino decays will decay through the channel $\tilde{W}^\pm \to \pi^\pm + \tilde{W}^0$. This will give rise to signals analogous to the previous case, with the additional feature of charged disappearing tracks from the long-lived $\tilde{W}^\pm$~\cite{distracks}. 
\item  One or both gluinos decay to charged winos and $\ell_{\tilde{W}^\pm \to \pi^\pm + \tilde{W}^0} \gg \ell_{\tilde{W}^\pm \to {W}^\pm + \tilde{a}}$.  In this regime the majority of charged winos produced from gluino decays will decay as $\tilde{W}^\pm \to W^\pm + \tilde{a}$, with a displacement $100~\mu{\rm m} < \ell_{\tilde{W}^\pm \to {W}^\pm + \tilde{a}} \ll 6$ cm.  If the gluinos both decayed to charged winos the collider signature is $jjjj+W^\pm W^\pm + \text{MET}$, with a modest displacement of both $W$ vertices, and potentially also displacement of the jet pair vertices.  If the gluino decayed to one charged and one neutral wino the signature would be  $jjjj+W^\pm + \gamma/Z + \text{MET}$ and again a modest displacement of the gauge boson vertices.
\end{itemize}

\section{Strong CP problem}
\label{seccp}
The theory we presented predicts that the CP-violating $\theta$ parameter of QCD is a number of order unity. This happens because the relaxion is stabilised at one of the local minima of the potential in \eq{pinpot}, and thus it is necessarily displaced by an amount of order one from the minima of the periodic potential. This result is in blatant contradiction with experiments, since the limit on the neutron electric dipole moment implies $|\theta |<10^{-10}$.  The problem is severe because any hypothetical solution setting $\theta =0$ at high energy would be undone by the relaxation mechanism operating at low energies. This difficulty is endemic in all relaxation mechanisms that employ the QCD axion as driving agent~\cite{GKR}. Of course, any model that claims to be realistic must solve this problem. Here we will only sketch some ideas on how to address the issue, but we will not attempt to construct a complete model, leaving this task to future work. We will present three possible ways to tackle the problem (two of them are adaptations to the supersymmetric case of the solutions suggested in ref.~\cite{GKR}). Each solution has some drawbacks.

\subsection{Inflaton-dependent relaxion potential}
The first class of solutions employs the idea that the PQ-breaking potential that drives the evolution of the relaxion towards electroweak breaking can be present during inflation, but disappear after reheating. In this way, after the end of inflation, the relaxion will be able to rearrange itself very close to a minimum of the periodic potential, driving $\theta$ towards zero with an unsubstantial change of the Higgs mass. 

In our context, the idea can be realised by adding to the superpotential a small coupling $\lambda$ between the inflaton ($I$) and relaxion ($S$) chiral superfields
\beq
W=\left( m-\lambda I \right) \frac{f^2\, S^2}{2} + \frac{m_I \, I^2}{2} \, .
\eeq
Here, just for illustration, we have taken a simple mass term for the inflaton while we choose a canonical K\"ahler potential. Our dynamical assumption is that, during inflation, the scalar component of $I$ (called $\varphi_I$) and the relaxion ($a$) are displaced from their minima, while the srelaxion sits at the vacuum $s=0$. We find that, on the inflaton background, the effective relaxion mass (defined as $V(a) = m_{\rm eff}^2 f^2 a^2/2$) and auxiliary field are given by
\bea
m_{\rm eff}^2(\varphi_I ) &=& (m-\lambda \varphi_I )^2 +\lambda m_I \varphi_I \, , \\
F_{\rm eff}(\varphi_I ) &=& i(m+\lambda \varphi_I )\, \frac{a}{\sqrt{2}} \, .
\eea

Let us suppose that, during inflation, the inflaton lies somewhere in the range $m^2/m_I\ll \lambda \varphi_I \ll m$. Then, the supersymmetry-breaking scale during and after inflation is always about the same, $ F_{\rm eff}(\varphi_I )\approx F_{\rm eff}(0 )$. On the other hand, the relaxion mass during inflation, $m_{\rm eff}^2(\varphi_I )\approx \lambda m_I \varphi_I $ is much larger than its value today $m_{\rm eff}^2(0 )\approx m^2$. The ratio of these two masses squared corresponds to the reduction factor in the value of $\theta$. Thus, we obtain
\beq
\theta \approx  \frac{m_{\rm eff}^2(0 )}{m_{\rm eff}^2(\varphi_I )} \ll 1 \, .
\eeq
We must also require the condition $\lambda < m^2/(\mu_0 f)$ to ensure that a quartic term in the potential $V(a)$ gives a negligible contribution with respect to the mass term. In conclusion, this mechanism can efficiently reduce $\theta$ without affecting the scanning of the supersymmetry-breaking scale, nor reducing its value after inflation.

However, the inflationary conditions discussed in section~\ref{secinf} become now very constraining. The requirement that the inflaton dominates the vacuum energy, {\it i.e.} the analogue of \eq{pif3}, gives
\beq
H > \left( \frac{f}{10^9~\gev}\right) \left( \frac{\mu_0}{10^5~\gev}\right) \left( \frac{10^{-10}}{\theta}\right)^{1/2}\gev
 \, .
\label{pifniente}
\eeq
For sparticle masses in the tens of TeV range, one can barely satisfy the requirement $H< \Lambda$, see \eq{pif5}. On the other hand, the condition for classical evolution is not satisfied and it remains dubious if the relaxion can have a significant probability of reaching the correct vacuum.

\subsection{Inflaton-dependent instanton barriers}
An alternative strategy to deal with the strong-CP problem is to have the field evolution to occur during a period of inflation in which the Hubble constant exceeds the QCD strong coupling scale, $H > \Lambda$.  Previously this regime has been avoided as the instanton effects, and hence the axion potential, are exponentially suppressed.  However, we will see that this may be advantageous.

\begin{figure}[t]
\centering
\includegraphics[height=2.9in]{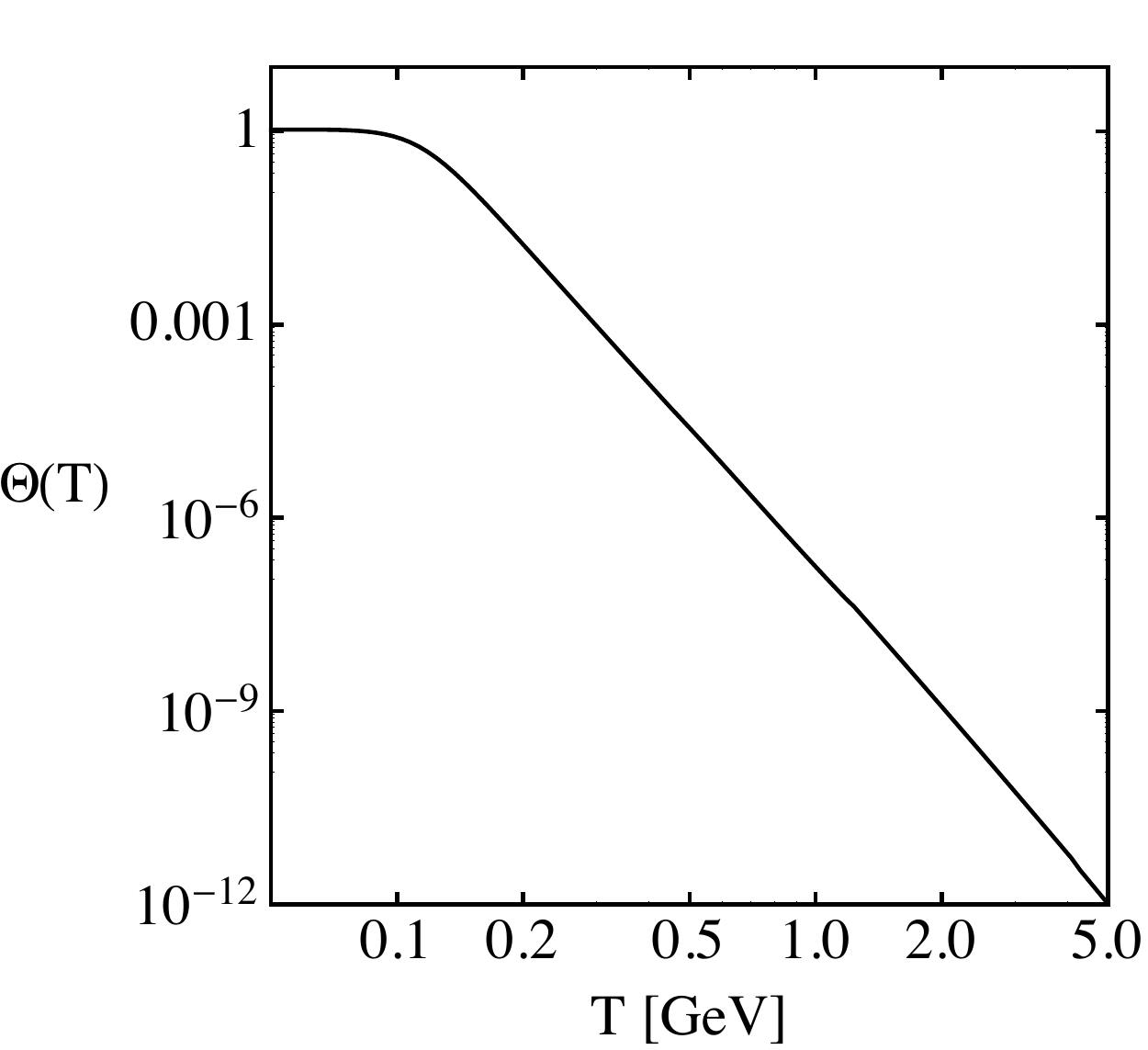}
\caption{The suppression of the axion potential at finite temperature, calculated using the approximate expression from ref.~\cite{Wantz:2009it}.}
\label{fig:suppaxiont}
\end{figure}

Let us first consider the behaviour of a thermal system, as this will lead us to an interesting analogy.
At temperatures well below $\Lambda$ the instanton effects are unsuppressed and the effective axion potential may be calculated from the chiral Lagrangian.  The resulting potential is given by
\begin{equation}
V_a \sim  \Lambda^4 \cos a ~~.
\end{equation}
At temperatures far above $\Lambda$ the potential may be calculated in perturbation theory and, to lowest order is given by $V_a \sim T^4 \exp [-{2 \pi}/{\alpha_s (T)}]$ \cite{Gross:1980br}.  As it is not possible to estimate the potential in the cross-over regime where $T\sim \Lambda$, we make a simple parametrisation
\begin{equation}
V_a (T) \sim  \Lambda^4\,  \Theta (T) \cos a \, ,
\label{eq:effT}
\end{equation}
where the function $\Theta (T)$ encodes the finite temperature suppression.  This suppression has been estimated in a number of different temperature regimes (see e.g. \cite{Gross:1980br,Turner:1985si}).  In \fig{fig:suppaxiont} we plot the finite temperature suppression factor as calculated more recently, by using the approximate expressions in \cite{Wantz:2009it}.\footnote{We flip the sign of the $d_0^{(4)}$ coefficient given in \cite{Wantz:2009it}, otherwise the complete function is not continuous.}

Rather than considering bone-fide finite temperature effects, let us instead consider inflationary Hubble scales which exceed $\Lambda$.  There are two different perspectives for understanding how, for $H > \Lambda$, the axion potential is suppressed. The first is that instanton effects will only be physical for instanton sizes which are contained within the horizon, {\it i.e.} only instantons of radius $\rho < 1/H$ will contribute to the path integral.  
The path integral is IR divergent and a cutoff is needed, with the integration over the radius extending up to distances $\rho < 1/\Lambda$.  However, if $1/H < 1/\Lambda$ then a new cosmological IR cutoff is imposed and the integration is limited to distances $\rho < 1/H$.
Since for modes of small size ($\rho < 1/H$) the gauge coupling is perturbative, the instanton effect is exponentially suppressed.

The second perspective is that de-Sitter space may be thought of as exhibiting a finite-horizon Gibbons-Hawking temperature $T_H \sim H/2 \pi$~\cite{hawk}.  While it should be kept in mind that this temperature is physically different from the usual interpretation in statistical mechanics, an estimate of the axion potential may be determined by substituting $T_H$ into \eq{eq:effT}.

\begin{figure}[t]
\centering
\includegraphics[height=1.8in]{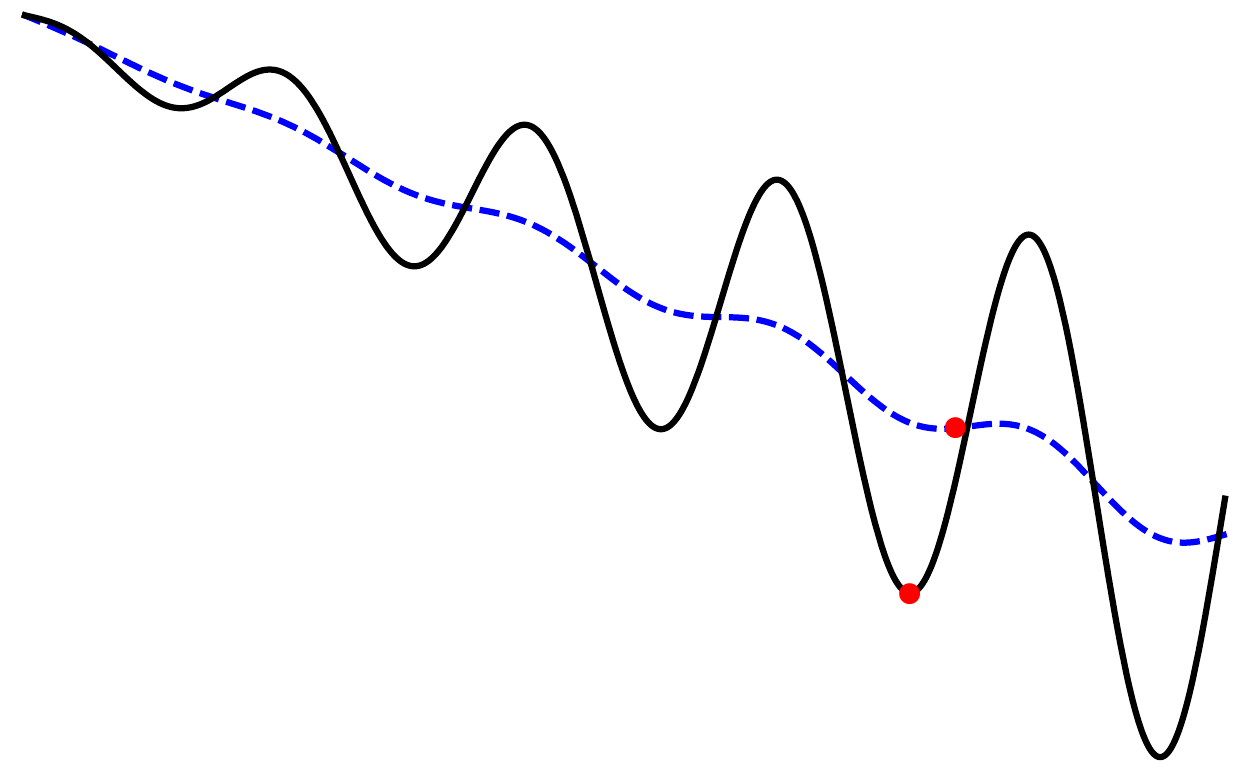}
\caption{A schematic illustration of the resolution of the relaxion strong-CP problem with inflaton-dependent instanton barriers.  During inflation the axion-like potential is suppressed (blue dashed line).  At late times the axion-like potential is unsuppressed and has grown relative to its value during inflation (solid black line).  This change shifts the final relaxion minimum closer to a value in which the effective $\theta$ is much smaller, as shown by the example red minima.}
\label{fig:superdiag}
\end{figure}

Let us now consider the relaxation during an epoch with $T_H>\Lambda$.  We will assume that the full evolution of the relaxion occurs during an inflationary period with constant $H$.
 
The effective potential for the relaxion during inflation is given by \eq{pinpot}, which now becomes
\begin{equation}
V(a) = \frac{m^2 f^2}{2} a^2 + \Lambda^4\,  \Theta (T_H) \cos a \, .
\end{equation}
The relaxion evolution of this model is identical to our original model, with the exception that the 
heights of the barriers in the instanton-induced potential are now suppressed.  For the relaxion to stop after electroweak breaking we need to satisfy
\begin{equation}
  \Lambda^4 \Theta (T_H)  \approx f^2 m \mu_0 \, .
\end{equation}

As before, the effective strong-CP angle $\theta$ at the end of relaxion stabilisation will be of order unity.  However, after inflation has ended the Hubble parameter will have dropped below the QCD scale and in this post-inflationary epoch the amplitude of the periodic potential will have grown, while the value of $m$ will remain the same as before.  This means that at late times the relaxion potential is dominated by the usual axion potential and the relaxion will evolve towards the new minimum of the potential, appearing almost identical to the QCD axion.  This would then solve the strong-CP problem.

Quantitatively, the relaxion at the minimum is such that $\sin a \approx m^2f^2a_*/(\Lambda^4 \Theta )$. This means that the effective value of $\theta$ changes from the inflationary epoch (when $\Theta$ must be evaluated at $T_H$) to late times (when $\Theta =1$) by a factor of $\Theta(T_H)$. Thus the strong CP angle today is $\theta \approx \Theta (T_H)$.

We see from \fig{fig:suppaxiont} that for $T_H \gtrsim 2.8$ GeV ($H \gtrsim 18$ GeV) we have $|\Theta (T_H)| \lesssim 10^{-10}$ and the strong-CP problem is resolved.  This mechanism is sketched in \fig{fig:superdiag}.

Unfortunately the condition for classical evolution is not satisfied once again.  We speculate that an alternative strategy would be to take $H < \Lambda$ as usual and suppress the axion potential during inflation by genuine finite temperature effects in the visible sector $T > \Lambda$.  This may be achievable by allowing the inflaton field to decay to visible sector fields throughout the inflationary period, in a setup similar in spirit to warm inflation \cite{Berera:1995ie}.  The details of such a scenario remain to be investigated.

\subsection{Non-QCD relaxion}
Another solution to the strong-CP problem may be found by realising the relaxion as the axion-like field of a new gauge group beyond the SM.  This follows the original proposal of ref.~\cite{GKR} and our task here is to describe the supersymmetric counterpart.
For our scenario this requires supersymmetrizing the non-QCD model of \cite{GKR}.  A new $\text{SU}(3)^\prime_{c}$ gauge group is introduced and vector-like matter in the fundamental and anti-fundamental of $\text{SU}(3)^\prime_{c}$ is added.  Some of the matter fields carry the electroweak quantum numbers of the left-handed lepton superfield, thus they are labelled $L$ and $L^c$, while the others are electroweak neutral superfields $N,N^c$.  The theory is described by the usual K\"ahler potential and a superpotential given by
\bea
W &=&   C_a(S) \, {\rm Tr} {\cal W}'_a{\cal W}'_a + M_L L L^c + M_N N N^c +y_u  H_u L N^c+y_d  H_d L^c N  ~~.
\label{laggnonQCD}
\eea
The electroweak charged fields $L,L^c$ must have masses at or above the weak scale to have evaded collider detection.  $\text{SU}(3)^\prime_{c}$ strong coupling leads to confinement of the matter fermions. We must require the condensation of these fields to be suppressed by taking $\Lambda' < M_L$, otherwise $\langle L L^c \rangle \sim {\Lambda'}^3$ and electroweak symmetry breaking would be dominated by a technicolour phase.  There are additional restrictions on the mass spectrum related to technical naturalness, which may be found in ref.~\cite{GKR}.

In order for the relaxation mechanism to work the mass of the lightest Dirac fermion charged under $\text{SU}(3)^\prime_{c}$ must be dominated by the Higgs vev.  This mass contribution is found after we integrate out $L,L^c$,
\bea
W &\sim&  y_u y_d \frac{H_u H_d}{M_L} N N^c  ~~,
\label{checifa}
\eea
from a supersymmetric seesaw.  In order for this mass term to dominate there is a form of $\mu$-problem associated with taking a small value for the superpotential parameter $M_N$. However this is technically natural so long as it is not more than a loop factor below $M_L$. Unfortunately supersymmetry, which is broken at a high scale, cannot protect $M_N$ further than that.

The dynamical evolution of this setup proceeds as before.  The pseudoscalar field $a$ contained in $S$ will roll down a shallow potential, scanning the supersymmetry breaking scale as it does so.  At some point, when $\langle h_{u,d} \rangle \neq 0$, a dominant mass contribution for the $N, N^c$ fields is induced by the interaction in \eq{checifa}.
Confinement due to the $\text{SU}(3)^\prime_{c}$ strong coupling generates an axion-like potential for $a$ which stops the field from rolling.  This creates a metastable minimum in which the supersymmetry breaking $F$-term has been stabilised at a large value and the Higgs vev at a small value.

A difference between this setup and the non-QCD model of \cite{GKR} lies in the couplings of the relaxion.  In our realisation $S$ is the source of supersymmetry breaking.  Thus to generate visible sector soft masses, including gaugino masses, $S$ must couple to the visible sector superfields, hence the non-QCD relaxion must have couplings to the visible sector gauge fields via the usual axion-like interactions $a G \widetilde{G}$, $a F \widetilde{F}$.  For this reason the non-QCD relaxion may be detectable through the usual axion experimental strategies.  As it obtains its dominant mass from an additional gauge group it would appear much like a QCD axion, albeit with mass that is anomalously large.  To resolve the strong-CP problem there must also exist the usual QCD axion and it must not couple to $\text{SU}(3)^\prime_{c}$ to enforce that its dominant mass contribution will come from QCD and not QCD$'$.  Hence both the QCD axion and the QCD$'$ relaxion may be detectable in this setup.\\

\section{Summary}

For the ease of the reader we summarise here our results.

Section~\ref{secframe} describes our theoretical framework, which is based on a supersymmetric effective theory valid below the PQ scale $f$, with SM superfields and a chiral superfield for the relaxion. To mimic the effect of monodromy we introduce an explicit breaking of the shift symmetry through a small mass term, which generates the relaxion potential. We show how the background value of the relaxion breaks supersymmetry and we compute the induced soft terms.

In section~\ref{secrelax} we explain how the evolution of the relaxion leads to electroweak breaking when the scanning soft terms become of the order of the supersymmetric higgsino mass parameter $\mu_0$. The back-reaction from QCD instantons stops the evolution of the relaxion and stabilises the scale of supersymmetry breaking at a value of order $\mu_0$.

Section~\ref{secinf} describes the conditions under which the inflationary dynamics is compatible with the relaxation mechanism. The most stringent constraints come from the conditions that {\it (i)} the relaxion does not  dominate the vacuum energy, so that it does not affect the dynamics of inflation, and {\it (ii)} the relaxion evolution is determined by the classical force and not by the quantum random walk, so that the slow-rolling field tracks the classical potential. The combination of these two conditions imply a strong upper bound on the Hubble rate during inflation and an upper bound on the scale of supersymmetry breaking of some hundreds of TeV. Once these two conditions are satisfied, it is guaranteed that {\it (i)} gravity-mediated effects on the soft terms from the inflaton sector are negligible and {\it (ii)} the relaxion is in the slow-roll regime. A further requirement is that inflation lasts for an astronomically large number of e-folds in order to give enough time to the relaxion to probe a sufficiently large portion of its shallow potential.

In section~\ref{secUV} we explore the limitations and the uncertainties associated with the super-Planckian excursion of the relaxion. The non-compact properties of the axion most probably require UV completions beyond the rules of quantum field theory and this prevents us from making definitive statements. However, we argue that such completion needs only to emerge in the Planckian domain and not necessarily at the lower scale $f$. Using an effective field-theory approach, we observe that selection rules could keep Planckian effects under control in the region of interest of the relaxion potential. Moreover, the robustness of the mechanism under modifications of the relaxion potential makes us more confident that relaxation may survive Planckian effects. On the other hand, we show that relaxation is inconsistent with the conjecture of gravity as the weakest force.
We also remark that supergravity could lead to the interesting situation of an unstable nearly-flat direction along which the relaxion could slide. This would give a natural explanation of the initial conditions at early times. The relaxion would start at typical Planckian values, then grow enormously driven by the dynamics of the runaway direction to be stopped only by QCD instantons. The vastly super-Planckian values of the relaxion would not result from an assumption on initial conditions, but rather from the dynamical evolution. It is not clear to us if this scenario can be made compatible with realistic mechanisms for the cancellation of the cosmological constant. 

In section~\ref{secstru} we elucidate the mechanism of supersymmetry breaking, which is particularly simple in terms of field content -- a single chiral superfield -- but complicated in terms of dynamics because of the simultaneous participation of vastly different energy scales, varying from hundreds of TeV to hundreds of MeV. Since all these scales play an active role in the process, we devise a simple but effective way to capture the relevant physics and obtain an effective theory of the interactions between the relaxion and the relaxino below the QCD scale. The supersymmetric relations among couplings and the non-perturbative QCD chiral condensate conspire to make the relaxino exactly massless at the metastable vacuum. This indicates that the relaxino must be identified with the Goldstino and supersymmetry is indeed spontaneously broken.

Once gravity is turned on, the relaxino plays the role of the spin-1/2 components of the gravitino. In our scenario, the relaxino is the LSP, since its mass is a factor $f/M_P$ smaller than the typical soft mass. The cosmology of the relaxino is discussed in section~\ref{secgravitino}. Thermal abundance considerations require that the reheating temperature after inflation should not be larger than the typical soft mass. When this bound is saturated, the relaxino could be the dark matter. More generally, dark matter could be made of two species, being a combination of relaxinos and relaxions.

In section~\ref{secph} we considered the low-energy relaxion phenomenology and high-energy gaugino phenomenology at colliders.  Although the scalar superpartners are likely to be out of reach at the LHC, the gaugino masses are suppressed by an additional loop factor and may be within LHC and possible future collider reach. So relaxation gives a realisation of Split Supersymmetry~\cite{split} free from the naturalness problem.  As the LSP is the relaxino (\ie\ gravitino), the model predicts gaugino NLSP decays that may be prompt, displaced, or even outside the detector, giving a variety of characteristic signals.  The specific collider signatures depend on the nature of the NSLP, and they typically involve jets, missing energy, possibly two or more displaced vertices, and additional electroweak gauge bosons.  If the NLSP decays outside the detector then a gluino NLSP would generate R-hadron signatures and a bino or wino NLSP would lead to a jets and missing energy signature, possibly accompanied by disappearing charged tracks.

In section~\ref{seccp} we investigate three scenarios to address the problematic strong-CP prediction of the relaxion model.  In the first scenario we sketch a supersymmetric inflaton-relaxion coupling in which the slope of the potential breaking the shift symmetry is generated during inflation but drops significantly afterwards.  This is an adaptation of a similar setup described in ref.~\cite{GKR}.  Unfortunately in the supersymmetric model it is challenging to satisfy the constraint that classical dynamics dominates the evolution of the relaxion field.  In the second scenario we considered a new possibility that the axion-like potential induced by QCD instantons may be suppressed during relaxation and grow afterwards to force the strong CP angle $\theta$ to small values.  For this scenario it also appears difficult to enforce classical evolution of the relaxion, although we speculate that modified scenarios involving a type of warm inflation may be a promising avenue for future investigation.  Finally, in the third scenario we adapt a model of ref.~\cite{GKR} where additional matter is added such that relaxation is a result of chiral symmetry breaking due to a non-QCD gauge group.  This scenario satisfies the classical evolution constraint.  

\subsubsection*{Acknowledgments}

We would like to thank Tevong You for collaborating in the initial stages of this work.  We gratefully acknowledge useful discussions with A.~Arvanitaki, N.~Craig, S.~Dimopoulos, P.~Draper, J.~March-Russell, A.~Pomarol, R.~Rattazzi, P.~Schwaller, F.~Zwirner. M.M. is grateful to N.~Craig and P.~Draper for conversations on SUSY implementations of the relaxion.
 
\appendix
  
\section{Computation of the soft terms}
\label{app:soft}
In this appendix we outline the computation, based on the method of ref.~\cite{wave}, of the soft terms coming from the Langrangian in \eq{lagg}, in presence of the supersymmetry-breaking background of the relaxion superfield $S={\tilde S}+\theta^2 F$. Here ${\tilde S}=(s+ia)/\sqrt{2}$ denotes the complex scalar component and $F$ is the auxiliary field. On this background, the functions appearing in the Lagrangian can be expressed as
\bea
Z_i (S+S^\dagger ) &=& Z +Z'(F\theta^2+F^*{\bar \theta}^2)+Z''|F|^2\theta^4\, , \label{pin1}\\
U(S+S^\dagger ) &=&U'F^*{\bar \theta}^2+U''|F|^2 \theta^4 \, , \label{pin2}\\
e^{-qS} &=& e^{-q{\tilde S}}(1-qF\theta^2) \, . \label{pin3}
\eea
On the right-hand side of eqs.~(\ref{pin1})--(\ref{pin3}), $Z$ and $U$ are functions of the variable ${\tilde S}+{\tilde S}^\dagger = \sqrt{2}s$ and primes denote derivatives with respect to this variable.

To obtain physical masses, we need to work in a basis in which the kinetic terms for the SM-sector fields are canonically normalised. This is achieved by defining the rescaled chiral superfields
\beq
\hat \Phi_i = Z_i^{1/2} \left[ 1 + (\ln Z_i)' F\theta^2 \right] \Phi_i \, .
\eeq
In terms of the rescaled superfields, the Lagrangian in \eq{lagg} becomes
\bea
\Lag &=& \int d^4 \theta \, \Big[ 1+(\ln Z_i)'' |F|^2\theta^4\Big] \, {\hat \Phi}_i^\dagger {\hat \Phi}_i \nonumber \\
&+& \left( \int d^4 \theta \, \frac{e^{-q{\tilde S}}}{( Z_{H_u}Z_{H_d})^{1/2}}\, \Big[ U' F^*{\bar \theta}^2+\left( U'' -Q U'\right) |F|^2\theta^4\Big]\,  {\hat H}_u {\hat H}_d \right.  \nonumber \\ 
&+& \left. \int d^2 \theta \,  \frac{\mu_0 \, e^{-q{\tilde S}}}{( Z_{H_u}Z_{H_d})^{1/2}} \left( 1 -Q F\theta^2\right) \, {\hat H}_u {\hat H}_d 
\right.  \nonumber \\ 
&+& \left. \int d^2 \theta \, Y_{ijk} \left( 1-P_{ijk} F\theta^2 \right) {\hat \Phi}_i {\hat \Phi}_j {\hat \Phi}_k \right. \nonumber \\
&+&  \left. \int d^2 \theta \,\left\{ C_a(S)+\sum_i \frac{T_a^i}{16\pi^2}\Big[\ln Z_i+(\ln Z_i)' F \theta^2 \Big]\right\}{\rm tr}\,{\cal W}_a{\cal W}_a+
 \hc \right)
 \label{pinto}
\eea
\bea
Q&\equiv& q+(\ln Z_{H_u})' + (\ln Z_{H_d})'  \, , \\
P_{ijk}&\equiv& (\ln Z_i)' + (\ln Z_j)' + (\ln Z_k)' \, .
\eea
Here $Y_{ijk}$ are the running Yukawa couplings, which include the wave-function renormalisation.
We define $T^i_a$ as the Dynkin index of the $\Phi_i$ representation under the gauge group $a$ ($T^i =1/2$ or  $T^i =N$ for a fundamental or an adjoint of SU($N$), respectively).

From \eq{pinto} we can immediately read off the soft terms
\bea
{\tilde m}_i^2 &=&-(\ln Z_i)'' |F|^2\, , \label{ciop1}\\
A_{ijk} &=& Y_{ijk}\, \big[  (\ln Z_i)' + (\ln Z_j)' + (\ln Z_k)' \big] \, F \, ,\\
\mu &=& \frac{e^{-q{\tilde S}}}{( Z_{H_u}Z_{H_d})^{1/2}}\, \left( \mu_0 + U'F^* \right) \, , \label{ciop2}\\
B_\mu &=& \big[ q+(\ln Z_{H_u})' + (\ln Z_{H_d})'  \big] \, F \mu -   \frac{ e^{-q{\tilde S}}U''|F|^2}{( Z_{H_u}Z_{H_d})^{1/2}} \,  ,\label{ciop3} \\
M_{{\tilde g}_a}&=& \frac{\alpha_a}{4 \pi}  \Big[ c_a -\sum_i T_a^i (\ln Z_i)' \Big] \, F+
 \frac{\alpha_a\, B_\mu}{2\pi\, \mu}\, f\left( \frac{\mu^2}{m_H^2}\right) 
(\delta_{a1}+\delta_{a2})\, , \label{ciop4}
\eea
with $f(x)=(x\ln x)/(x-1)$. 

Note that the second term in \eq{ciop4} is a gauge-mediation effect from the Higgs superfields $H_{u,d}$. It cannot be neglected here because it is parametrically comparable with the first term.  In \eq{ciop4}, $m_H$ is the heavy Higgs mass  defined in \eq{defmh}.
 
It is useful to remark that the value of $q$ in the Lagrangian in \eq{lagg} depends on the field basis. Let us consider the $S$-dependent superfield redefinition
\beq
\Phi_i \to e^{q_i S} \, \Phi_i \, ,
\label{flip}
\eeq
where $q_i$ are the corresponding PQ charges. After this transformation, 
the Lagrangian is obtained from \eq{lagg} with the replacements 
\beq
q\to q-q_{H_u}-q_{H_d} \, , ~~~ Z_i \to e^{q_i(S+S^\dagger )}\, Z_i \, , ~~~c_a \to c_a + \sum_i T^i_a q_i \, ,
\label{trans}
\eeq
 The variation of $c_a$ in \eq{trans} is induced by the quantum anomaly of the PQ symmetry. The Yukawa interactions remain invariant because we are assuming that they respect PQ. 
 
With the transformation in \eq{flip} one can eliminate the $e^{-qS}$ factor from the superpotential, thus exhibiting the basis dependence of the value of $q$. Note that the soft terms in eqs.~(\ref{ciop1})--(\ref{ciop4}) are manifestly invariant\footnote{Under the transformation in \eq{trans}, both $\mu$ and $B_\mu$ change by an overall phase $\exp[i(q_{H_u}+q_{H_d})a/\sqrt{2}]$. However, this phase is irrelevant since physical quantities can depend only on the basis-independent combination arg$(m_\lambda \mu B_\mu^*)$.} under the transformation in \eq{trans} and therefore are independent of the field basis, as physical quantities should.

\section{Conditions for electroweak breaking}
In this appendix we derive the conditions under which the electroweak symmetry is broken by the dynamical evolution of the relaxion, as it rolls down its potential. We parametrize the soft-breaking parameters as
\beq
m_{H_u}^2 =c_u \, m^2a^2 \, ,~~~
m_{H_d}^2 =c_d \, m^2a^2 \, ,~~~
\mu =\mu_0- c_\mu \, ma \, ,~~~
B_\mu=c_0\, \mu \, ma+c_B  \, m^2a^2 \, ,
\label{coeffi}
\eeq
where $c_i$ are model-dependent coefficients that we take to be independent of $a$ and of order unity. For simplicity, we take all $c_i$ real and we can choose $c_d$, $c_B$, and $\mu_0$ positive. The corresponding coefficients $c_i$ for squarks and leptons are taken such that the vacuum does not spontaneously break colour or electric charge.

The order parameter of electroweak breaking is the determinant of the Higgs mass matrix, given by \eq{mscr}. Using \eq{coeffi}, we find
\beq
{\mathscr D}(a) 
= m^4 a^4 \left[ \left( \frac{\mu_0}{ma} -c_\mu \right)^4 +\left( c_u+c_d-c_0^2\right) \left( \frac{\mu_0}{ma} -c_\mu \right)^2 -2c_0c_B \left( \frac{\mu_0}{ma} -c_\mu \right) +c_uc_d -c_B^2 \right] \, . 
\eeq
We require that, during the initial stage of the relaxion evolution, electroweak symmetry is preserved (${\mathscr D}(a) >0$ for $a\gg \mu_0 /m$). This implies
\beq
\left(c_u+c_\mu^2\right)\left(c_d+c_\mu^2\right) > \left(c_B-c_0c_\mu \right)^2 ~~~~~\hbox{(no EW breaking at large $a$)}\, .
\label{cond1}
\eeq
As long as ${\mathscr D}(a) >0$, the condition for stability of the Higgs potential along the $D$-flat direction ($m_{H_u}^2+m_{H_d}^2+2\mu^2 > 2|B_\mu |$) is automatically satisfied.

As the relaxion rolls down its potential, ${\mathscr D}(a)$ decreases. However, an overall rescaling of ${\mathscr D}(a)$ does not trigger electroweak breaking. Instead, we want ${\mathscr D}(a)$ to change sign during the evolution of $a$, and thus the condition for approaching electroweak breaking is $d({\mathscr D}(a)/a^4)/da >0$. Imposing this condition at $a\gg \mu_0 /m$ implies
\beq
c_\mu \left( 2c_\mu^2+c_u+c_d-c_0^2\right) +c_0c_B >0 ~~~~~\hbox{(approach towards EW breaking)}\, .
\label{cond2}
\eeq
If $d({\mathscr D}(a)/a^4)/da$ remains positive as the relaxion rolls down its potential, eventually at a value $a=a_*$ the critical condition ${\mathscr D}(a_*)=0$ is achieved. The value of $a_*$ can be written as in \eq{astar},
where $c_*$ is a function of the coefficients $c_i$ in \eq{coeffi}, which is expected to be of order unity.

Although we cannot give a general analytic expression of $c_*$, we can easily compute it in two simple, but representative, cases. The first case is $c_0 =0$, in which electroweak breaking is achieved when $c_B^2>c_uc_d$, together with the conditions in eqs.~(\ref{cond1})--(\ref{cond2}), and $c_*$ is given by
\beq
c_*=\sqrt{2}\left[ \sqrt{2} c_\mu -\left( \sqrt{(c_u-c_d)^2+4c_B^2} -c_u-c_d \right)^{1/2}\right]^{-1} \, .
\eeq
The condition for the stability of the potential along the $D$-flat direction for any value of $a$ is $c_u+c_d>2|c_B|$.

The second case is $c_B=0$. Electroweak breaking is achieved when eqs.~(\ref{cond1})--(\ref{cond2}) are supplemented by the condition $c_0^2 >c_u+c_d+2\sqrt{c_uc_d}$, and $c_*$ is given by
\beq
c_*=\sqrt{2}\left[ \sqrt{2} c_\mu -\left( \sqrt{(c_u-c_d)^2+c_0^4-2c_0^2(c_u+c_d)}+c_0^2-c_u-c_d \right)^{1/2}\right]^{-1} \, .
\eeq
The condition for the stability of the potential along the $D$-flat direction for any value of $a$ is $c_u+c_d>2|c_0|c_\mu +c_0^2/2$. These two examples illustrate how it is always possible to find a range of parameters in which the relaxion evolution is driven towards the critical condition for electroweak breaking.

As discussed in section~\ref{secrelax}, once electroweak symmetry is broken, the relaxion is trapped by QCD instanton effects and its evolution stops. Nevertheless, it is interesting to study the relaxion potential for $a<a_*$, even if this range is not explored by the dynamical evolution. 

In the case $c_0=0$, as we decrease $a$ below $a_*$, we find that ${\mathscr D}(a)$, after exploring negative values, flips sign again and turns back positive at $a=a_{**}$ with
\beq
a_{**}=\frac{\sqrt{2}\mu_0}{m}\, \left[ \sqrt{2} c_\mu +\left( \sqrt{(c_u-c_d)^2+4c_B^2} -c_u-c_d \right)^{1/2}\right]^{-1} \, .
\eeq
This means that, as we decrease $a$, the potential barriers, besides being modulated by the decreasing value of the supersymmetry-breaking scale, completely disappear for $a<a_{**}$.

An even more complicated pattern is found in the case $c_B=0$. After becoming negative at $a=a_*$, ${\mathscr D}(a)$ flips sign first at $a=a_{**}$, then at $a=a_{-}$, and eventually turns back positive for $a<a_{+}$, where
\beq
a_{**}=\frac{\sqrt{2}\mu_0}{m}\, \left[ \sqrt{2} c_\mu -\left( -\sqrt{(c_u-c_d)^2+c_0^4-2c_0^2(c_u+c_d)}+c_0^2-c_u-c_d \right)^{1/2}\right]^{-1} \, ,
\eeq
\beq
a_{\pm}=\frac{\sqrt{2}\mu_0}{m}\, \left[ \sqrt{2} c_\mu +\left( \pm\sqrt{(c_u-c_d)^2+c_0^4-2c_0^2(c_u+c_d)}+c_0^2-c_u-c_d \right)^{1/2}\right]^{-1} \, .
\eeq

\end{document}